\documentclass[draft]{agujournal}

\draftfalse

\renewcommand{\textwidth}{15cm}
\usepackage{amsmath}  
\usepackage{hyperref}

\journalname{JGR-Planets}

\newcommand{\mitbf}[1]{
  \hbox{\mathversion{bold}$#1$}}

\begin{document}

%
%


\title{The influence of a fluid core and a solid inner core on the Cassini sate of Mercury}

%
%




\authors{Mathieu Dumberry \affil{1}}
 
\affiliation{1}{Department of Physics, University of Alberta, Edmonton, Alberta, Canada.}






\correspondingauthor{Mathieu Dumberry}{dumberry@ualberta.ca}




\begin{keypoints}
\item The Cassini state obliquity of Mercury's mantle spin axis deviates from that of a rigid planet by no more than 0.01 arcmin. 
\item For a core magnetic field above 0.3 mT, electromagnetic coupling locks the fluid and solid cores into a common precession motion. 
\item The larger the inner core is, the more the obliquity of the polar moment of inertia approaches that expected for a rigid planet.
\end{keypoints}

%
%

\begin{abstract}
We present a model of the Cassini state of Mercury that comprises an inner core, a fluid core and a mantle.  Our model includes inertial and gravitational torques between interior regions, and viscous and electromagnetic (EM) coupling at the boundaries of the fluid core.   We show that the coupling between Mercury's interior regions is sufficiently strong that the obliquity of the mantle spin axis deviates from that of a rigid planet by no more than 0.01 arcmin.  The mantle obliquity decreases with increasing inner core size, but the change between a large and no inner core is limited to 0.015 arcmin.  EM coupling is stronger than viscous coupling at the inner core boundary and, if the core magnetic field strength is above 0.3 mT, locks the fluid and solid cores into a common precession motion.  Because of the strong gravitational coupling between the mantle and inner core, the larger the inner core is, the more this co-precessing core is brought into an alignment with the mantle, and the more the obliquity of the polar moment of inertia approaches that expected for a rigid planet. The misalignment between the polar moment of inertia and mantle spin axis increases with inner core size, but is limited to 0.007 arcmin. Our results imply that the measured obliquities of the mantle spin axis and polar moment of inertia should coincide at the present-day level of measurement errors, and cannot be distinguished from the obliquity of a rigid planet.
\end{abstract}

\noindent{\bf Plain language summary:} The plane of Mercury's orbit around the Sun is slowly precessing about an axis fixed in space.  This entrains a precession of the spin axis of Mercury at the same rate, an equilibrium known as a Cassini state.  The angle between the spin axis and the normal to the orbital plane is known as the obliquity and remains fixed.  Observations have confirmed that Mercury's obliquity matches, within measurement errors, the theoretical prediction based on an entirely rigid planet.  However, we know that Mercury has a large metallic core which is liquid, although the central part may be solid.  In this work, we investigate how the presence of a fluid and solid core affect the Cassini state of Mercury.  We show that the internal coupling between the solid core, fluid core and the mantle is sufficiently strong that the obliquity of the mantle does not depart from that of a rigid planet by more than 0.01 arcmin, an offset smaller than the present-day error in measurements.  We also show that the larger the solid inner core is, the more the planet behaves as if it were precessing as an entirely rigid body.

\section{Introduction}

\begin{figure}
\begin{center}
    \includegraphics[height=10cm]{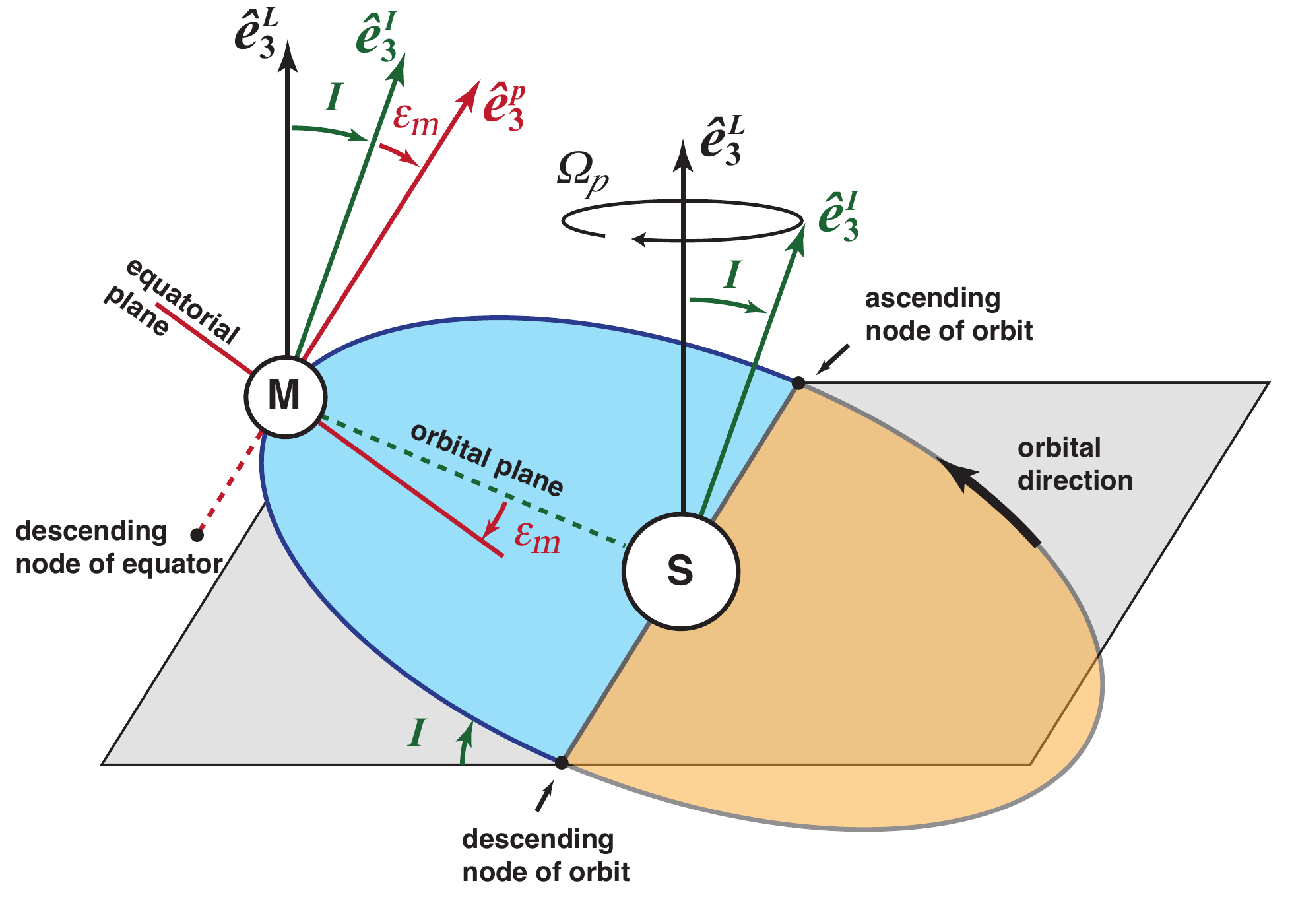} 
    \caption{\label{fig:cassini}  The orbit of Mercury (M) around Sun (S) with respect to the Laplace plane (grey shaded rectangle) and the Cassini state of Mercury.  The normal to the orbital plane ($\mitbf{\hat{e}_3^I}$) is offset from the normal to the Laplace plane ($\mitbf{\hat{e}_3^L}$) by an angle $I=8.5330^\circ$.  The symmetry axis of the mantle $\mitbf{\hat{e}_3^p}$ is offset from $\mitbf{\hat{e}_3^I}$ by $\varepsilon_m\approx2$ arcmin.  $\mitbf{\hat{e}_3^I}$ and $\mitbf{\hat{e}_3^p}$ are coplanar with, and precess about, $\mitbf{\hat{e}_3^L}$ in a retrograde direction at frequency $\Omega_p = 2\pi/325,513$ yr$^{-1}$.  The blue (orange) shaded region indicates the portion of the orbit when Mercury is above (below) the Laplace plane.  Angles are not drawn to scale.}
\end{center}
\end{figure}

Mercury is expected to be in a Cassini state  (Figure \ref{fig:cassini}) whereby its orbit normal and spin-symmetry axis are both coplanar with, and precess about, the normal to the Laplace plane \cite[][]{colombo66,peale69,peale06}. The orientation of the Laplace plane varies on long timescales, but its present-day orientation can be reconstructed from ephemerides data \cite[][]{yseboodt06,baland17}. Likewise, the rate of precession is also not observed directly, but is reconstructed by ephemerides data.  The latest estimate is a retrograde precession period of 325,513 yr with an inclination angle of $I=8.5330^\circ$ between the orbit and Laplace plane normals \cite[][]{baland17}.  Measurements of the obliquity $\varepsilon_m$, defined as the angle of misalignment between the spin-symmetry axis and the orbit normal, have been obtained by different techniques, including ground based radar observations \cite[][]{margot07,margot12}, and stereo digital terrain images \cite[][]{stark15} and radio tracking data \cite[][]{mazarico14,verma16,genova19,konopliv20} from the MErcury Surface Space ENvironment GEochemistry and Ranging (MESSENGER) spacecraft.   Within measurement errors, all techniques yield an obliquity which is coplanar with the orbit and Laplace plane normals and consistent with a Cassini state.  Furthermore, the observed obliquity angle ($2.042\pm0.08$ arcmin \cite[][]{margot12}, $2.029 \pm 0.085$ arcmin \cite[][]{stark15} and $1.968 \pm0.027$  \cite[][]{genova19} to list a few) matches that expected if Mercury occupies Cassini state 1.
 
The prediction of Mercury's obliquity is based on the assumption that the whole planet precesses as a single body.  However, we know that Mercury has a fluid core from two main lines of evidence.  First, Mercury's large scale magnetic field is intrinsic, and must be maintained by dynamo action \cite[][]{anderson11,anderson12,johnson12}.  This requires fluid motion in its metallic core, and hence that Mercury's core is at least partially liquid.  Second, the observed amplitude of the 88-day longitudinal libration is approximately twice as large as that expected if Mercury were librating as a rigid body \cite[][]{margot07,margot12,stark15}.  This indicates that it is only the mantle that librates, and that the outer part of the core is fluid.  These evidences do not necessarily imply that the whole of Mercury's core is fluid, but only that its outermost part must be.  A solid inner core may have nucleated at the centre although its size is not well constrained.  Inner core growth leads to planetary contraction, and the inferred radial contraction of $\sim7$ km since the late heavy bombardment \cite[][]{byrne14} places an approximate limit of 800 km on the inner core radius \cite[][]{grott11}.  However, the inner core could be larger if a significant fraction of its growth occurred earlier in Mercury's history.

With a fluid core, and possibly a solid inner core, the observed obliquity $\varepsilon_m$ reflects the orientation of the spin-symmetry axis of the precessing mantle and crust alone.  Neglecting dissipation, and at equilibrium in the Cassini state, the spin axis of the fluid core and the spin-symmetry axis of the inner core should both also precess about the normal to the Laplace plane in a retrograde direction with a period of 325,513 yr.  Both of these axes should also lie in the plane that defines the equilibrium Cassini state \cite[e.g.][]{dumberry16}, although their obliquity angles may be different than $\varepsilon_m$.  Whether the spin axis of the fluid core is brought into an alignment with the mantle obliquity depends primarily on the pressure torque (also referred to as the inertial torque) exerted by the centrifugal force of the rotating fluid core on the misaligned elliptical shape of the core-mantle boundary (CMB) \cite[][]{poincare10}. The more flattened the CMB is, the stronger the pressure torque is, and the more the fluid core is entrained into a co-precession at a similar obliquity to that of  the mantle.  The flattening of Mercury's CMB is not known.  But if one assumes that the topography of the CMB coincides with an equipotential surface at hydrostatic equilibrium with the imposed frozen-in mass anomalies in the upper mantle and crust, then the pressure torque at the CMB is sufficient to bring the fluid core into a close alignment with  the mantle \cite[][]{peale14}.  The spin axis of the fluid core is not expected to be exactly aligned with the spin-symmetry axis of the mantle, but sufficiently close that the resulting mantle obliquity does not differ much from that of a single body planet. Furthermore, viscous and electromagnetic (EM) coupling at the CMB can further restrict the misalignment between the mantle and core \cite[][]{peale14}.

If an inner core is present, its obliquity angle is determined by the sum of the torques acting on it.  This includes the gravitational torque from the Sun acting on its tilted figure, analogous to the torque applied on the tilted mantle that sets the obliquity $\varepsilon_m$.  In addition, the tilt of the inner core also depends on the gravitational torque imposed by the mantle and the pressure torque at the inner core boundary (ICB) imposed by the fluid core.  If the mantle gravitational torque dominates, the inner core tilt is expected to remain closely aligned with the mantle.  Conversely, if the pressure torque at the ICB is the largest, the inner core should instead be closely aligned with the spin axis of the fluid core.  A strong viscous and/or EM coupling at the ICB should also enforce a closer alignment between the rotation vectors of the inner core and fluid core.

It is on the basis of the observed mantle obliquity that the polar moment of inertia of Mercury is inferred \cite[e.g.][]{peale76,margot18}.  Inherent in this calculation is the built-in assumption that the mantle obliquity does not deviate from that of a rigid planet by a substantial amount.  However, the recent study by \cite{peale16} suggests that the inner core can be misaligned from the mantle by a few arcmin and that a large inner core can perturb the orientation of the spin vector of the mantle by as much as 0.1 arcmin.  This challenges the assumption that the observed obliquity reflects the orientation of the whole planet. 

Furthermore, if a large inner core is misaligned with the mantle, then the mantle spin axis does not coincide with the orientation of the polar moment of inertia of the whole planet.  This can introduce a systematic offset between different types of obliquity measurements.  Those based on tracking topographic features \cite[][]{margot07,margot12,stark15} capture the obliquity of the mantle spin axis.  While those based on the orientation of the gravity field \cite[][]{mazarico14,verma16,genova19,konopliv20} are instead tied to the orientation of the principal moment of inertia of the whole planet. An offset of the obliquity of the mantle spin axis with respect to the gravity field could be used to constrain the size of the inner core, even though this is difficult to do at present because the different estimates of the obliquity of the gravity field do not match well with one another.  

There is thus a significant interest in properly assessing how the presence of a solid inner core at the centre of Mercury may affect its Cassini state equilibrium.  Here, we present a model of Mercury's Cassini state that comprises a fluid core and solid inner core.  The model is an adaptation of a similar model developed to study the Cassini state of the Moon \cite[][]{dumberry16,stys18,organowski20}. The specific questions that motivate our study are the following.  First, we want to determine how large the misaligned obliquities of the fluid core and solid inner core can be and how they depend on model parameters.  Second, we want to assess by how much the mantle obliquity may differ from that of an entirely rigid Mercury, and third, by how much the obliquities of the spin-symmetry axis of the mantle and gravity field may differ.


\section{Theory}

\subsection{The interior structure of Mercury}
 
Our model of Mercury consists of four layers of uniform density: a solid inner core, a fluid outer core, a solid mantle, and a thin crust.   The outer radii of each of these layers, are denoted by $r_s$, $r_f$, $r_m$, and $R$, and their densities by $\rho_s$, $\rho_f$, $\rho_m$, and $\rho_c$, respectively. The inner core radius $r_s$ corresponds to the ICB radius, the fluid core radius $r_f$ to the CMB radius, and $R=2439.36$ km to the planetary radius of Mercury.  Compressibility effects from increasing pressure with depth are not negligible in the core of Mercury.  However adopting uniform densities simplifies the analytical expressions of the model while still capturing the first order rotational dynamics.  Uniform densities were also adopted by \cite{peale16} and following the same strategy facilitates comparisons between our results.

We build our interior model as detailed in \cite{peale16}. We first specify $r_s$, $\rho_s$ (or a density contrast at the ICB), the crustal density $\rho_c$ and crustal thickness $h=R-r_m$. The three unknowns $r_f$, $\rho_f$ and $\rho_m$ are then solved such that the interior model is consistent with the known mass $M$ and chosen values of the moments of inertia of the whole planet $C$ and that of the mantle and crust $C_m$.  

Each layer is triaxial in shape.   We denote the polar flattening (or geometrical ellipticity) by $\epsilon_i$, defined as the difference between the mean equatorial and polar radii, divided by the mean spherical radius.   Likewise, we denote the equatorial flattening by the variable $\xi_i$, defined as the difference between the maximum and minimum equatorial radii, divided by the mean spherical radius. As above, we use the subscript $i=s$, $f$, $m$ and $r$, to denote the polar or equatorial flattenings at the ICB, CMB, crust-mantle boundary (CrMB), and surface.

The measured polar and equatorial flattenings are taken from \cite{perry15} and their numerical values are given in Table \ref{tab:para}.  We then  assume that the ICB and CMB are both at hydrostatic equilibrium with the imposed gravitational potential induced by the flattenings at the CrMB and surface.  The flattenings at all interior boundaries are specified such that they are consistent with the observed degree 2 spherical harmonic coefficients of gravity $J_2$ and $C_{22}$; their numerical values are given in Table \ref{tab:para}.   Specifically, $J_2$ and $C_{22}$ are connected to the principal moments of inertia of Mercury ($C>B>A$) and to the polar and equatorial flattenings by  

\begin{subequations}
\begin{align}
J_2 & = \frac{C-\bar{A}}{M R^2} = \frac{8\pi}{15} \frac{1}{M R^2} \left[ (\rho_s-\rho_f) r_s^5 \epsilon_{s}  + (\rho_f-\rho_m) r_f^5 \epsilon_f + (\rho_m-\rho_c) r_m^5 \epsilon_m + \rho_c R^5 \epsilon_r  \right] \, , \label{eq:j2} \\
C_{22} & = \frac{B-A}{4 M R^2} =\frac{8\pi}{15} \frac{1}{4 M R^2} \left[ (\rho_s-\rho_f) r_s^5 \xi_s + (\rho_f-\rho_m) r_f^5 \xi_f + (\rho_m-\rho_c) r_m^5 \xi_m + \rho_c R^5 \xi_r  \right] \, . \label{eq:c22} 
\end{align}
\label{eq:j2c22}
\end{subequations} 
where $\bar{A}$ is the mean equatorial moment of inertia defined below.  The same procedure was used in \cite{peale16} and the mathematical details are given in Equations (18-20) of \cite{dumberry16} who adopted the same strategy in their interior modelling of the Moon.  Note that we neglect the misalignment between the triaxial shape of Mercury's surface topography and the axes of the principal moments of inertia, which amount to a polar offset of $\sim2^\circ$ and an equatorial offset of $\sim15^\circ$ \cite[][]{perry15}.

Once the densities and flattenings of all interior regions are known, we can specify the moments of inertia of the fluid core ($C_f>B_f>A_f$) and solid inner core ($C_s>B_s>A_s$) along with the mean equatorial moments of inertia

\begin{equation}
\bar{A} = \frac{1}{2} (A+B)  \, , \hspace*{0.5cm} \bar{A}_f = \frac{1}{2} (A_f+B_f)  \,,  \hspace*{0.5cm} \bar{A}_s = \frac{1}{2} (A_s+B_s)   \, . \label{ref:meanH}
\end{equation}
From these, we define the polar ($e$, $e_f$, $e_s$) and equatorial ($\gamma$, $\gamma_s$) dynamical ellipticities of the whole planet (no subscript), fluid core (subscript $f$) and solid inner core (subscript $s$), 
which enter our rotational model, 

\begin{subequations}
\begin{equation}
e = \frac{C-\bar{A}}{\bar{A}} \, \hspace*{0.5cm} e_f = \frac{C_f-\bar{A}_f}{\bar{A}_f} \, \hspace*{0.5cm} e_s = \frac{C_s-\bar{A}_s}{\bar{A}_s} \, ,
\label{eq:e}
\end{equation}
\begin{equation}
\gamma = \frac{B-A}{\bar{A}} \, \hspace*{0.5cm}  \gamma_s = \frac{B_s-A_s}{\bar{A}_s} \, .
\label{eq:e}
\end{equation}
\end{subequations}
We further note that $e$ and $\gamma$ are connected to $J_2$ and $C_{22}$ by

\begin{equation}
e = \frac{MR^2}{\bar{A}} J_2 \, ,\hspace*{0.5cm} \gamma = \frac{4MR^2}{\bar{A}} C_{22} \, .
\label{eq:egammaJ2C22}
\end{equation}

\begin{table}
\begin{tabular}{lll}
\hline
Mercury Parameter & Numerical value  & Reference \\ \hline
mean motion, $n$ &  $2\pi/87.96935$ day$^{-1}$  & \cite{stark15b} \\
rotation rate, $\Omega_o=1.5n$ & $2\pi/58.64623$ day$^{-1}$  & \cite{stark15b}\\
orbit precession rate, $\Omega_p$ & $2\pi /325,513$ yr$^{-1}$ & \cite{baland17} \\
Poincar\'e number, $\delta \omega = {\Omega_p}/{\Omega_o}$ & $4.9327 \times 10^{-7}$ & \\
orbital eccentricity, $e_c$ & $0.20563$ & \cite{baland17} \\
orbital inclination, $I$ & $8.5330^\circ$ & \cite{baland17} \\
mean planetary radius, $R$ & $2439.360$ km & \cite{perry15} \\
mass, $M$ & $3.3012 \times 10^{23}$ kg & \cite{genova19} \\
mean density, $\bar{\rho}$ & $5429.5$ kg m$^{-3}$ \\
$J_2$  & $5.0291 \times 10^{-5}$ &  \cite{genova19} \\
$C_{22}$ & $8.0415 \times 10^{-6}$ &  \cite{genova19} \\
polar surface flattening, $\epsilon_r$ & $6.7436 \times 10^{-4}$ &  \cite{perry15}\\
equatorial surface flattening, $\xi_r$ & $5.1243 \times 10^{-4}$ &  \cite{perry15}\\
\hline
\end{tabular}
\caption{\label{tab:para} Reference parameters for Mercury.  The mass $M$ is computed from $GM=22031.8636 \times 10^9$ m$^3$/s$^2$ taken from \cite{genova19}.  The mean density is calculated from $\frac{4\pi}{3} \bar{\rho}R^3 = M$.  The numerical values of $\epsilon_r$ and $\xi_r$ are calculated from $\epsilon_r = (\bar{a}-c)/R$ and $\xi_r = (a-b)/R$, where $\bar{a}=\frac{1}{2}(a+b)$ and where $a=2440.53$ km, $b=2439.28$ km and $c=2438.26$ km are the semimajor, intermediate and semiminor axes of the trixial ellipsoidal shape of Mercury taken from Table 2 of \cite{perry15}. $J_2$ and $C_{22}$ are computed from Equation (4) in the Supporting Information of \cite{genova19}.}
\end{table}

\subsection{The rotational model}

\begin{figure}
\begin{center}
    \includegraphics[height=7.5cm]{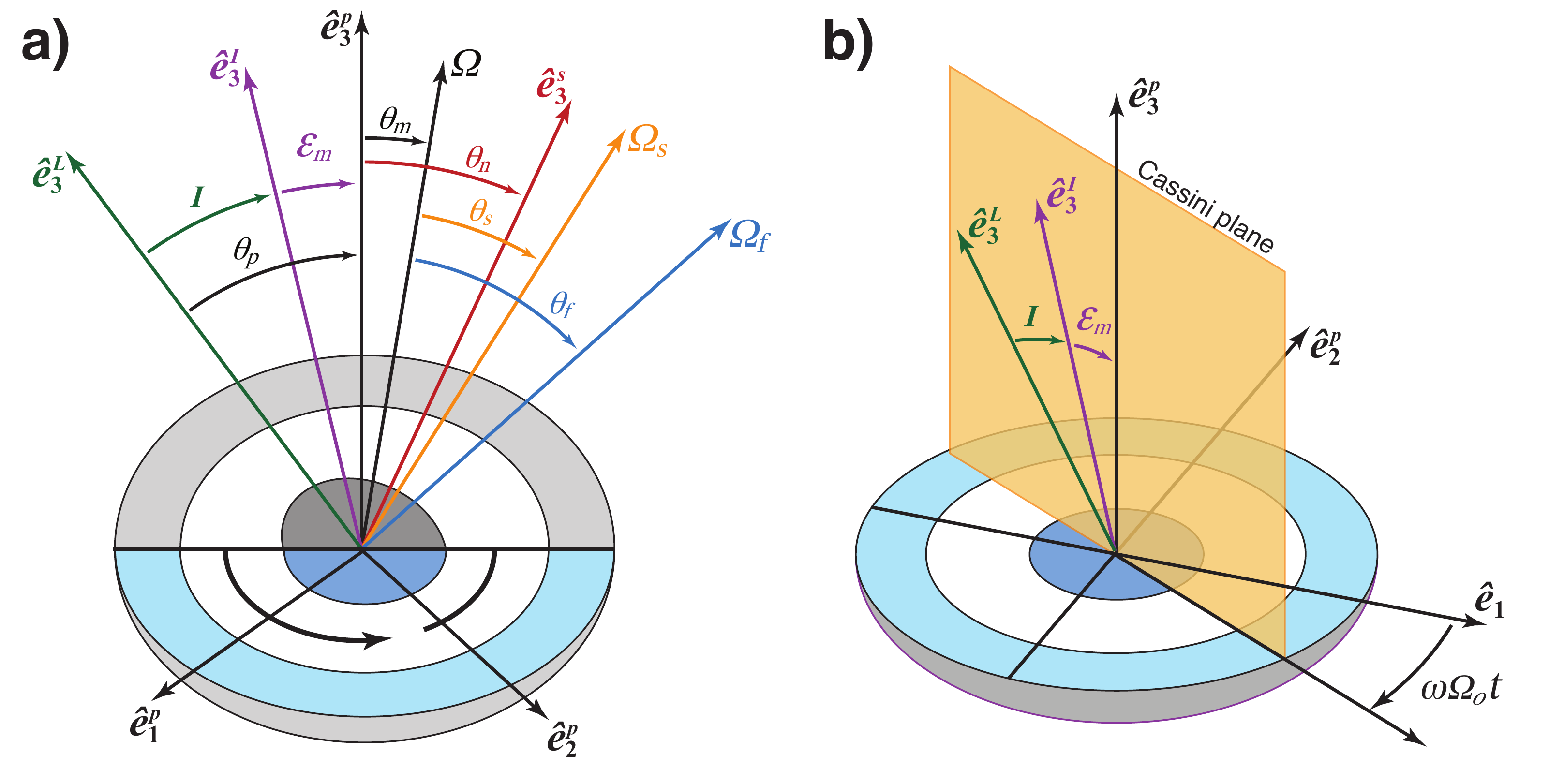} 
    \caption{\label{fig:angles}  Geometry of the Cassini state model of Mercury viewed (a) in the Cassini plane and (b) in a frame attached to the rotating mantle.   The orbit normal ($\mitbf{\hat{e}_3^I}$) is tilted by an angle $I=8.533^\circ$ from the Laplace normal ($\mitbf{\hat{e}_3^L}$) and the symmetry axis of Mercury's mantle ($\mitbf{\hat{e}_3^p}$) is tilted by an obliquity $\varepsilon_m$ with respect to $\mitbf{\hat{e}_3^I}$.  Shown in (a) are the orientations of the symmetry axis of the inner core ($\mitbf{\hat{e}^s_3}$), the rotation rate vectors of the mantle ($\mitbf{\Omega}$), fluid core ($\mitbf{\Omega_f}$) and inner core ($\mitbf{\Omega_f}$) and angles $\theta_p$, $\theta_n$, $\theta_m$, $\theta_f$ and $\theta_s$ in their Cassini state equilibrium.  All vectors and angles are in a common plane which we refer to as the Cassini plane.  The light grey, white, and dark grey ellipsoid represent a polar cross-section of the mantle, fluid core and inner core, respectively; blue shaded parts show an equatorial cross section.  The black curved arrow in the equatorial plane in (a) indicates the direction of rotation of the equatorial mantle axes $\mitbf{\hat{e}_1^p}$ and $\mitbf{\hat{e}_2^p}$ with respect to the Cassini plane.  Viewed in the frame attached to the rotating mantle (b), the Cassini plane is rotating at frequency $\omega \Omega_o = -\Omega_o - \Omega_p \cos I$ in the longitudinal direction.  The oblateness of all three regions and the amplitude of all angles are exaggerated for purpose of illustration.}
    \end{center}
\end{figure}

Mercury's rotation is characterized by a 3:2 spin-orbit resonance in which it completes 3 rotations around itself for every 2 orbital revolutions around the Sun. The orbital period is 87.96935 day and the sidereal rotation period is 58.64623 day \cite[][]{stark15b}.  These define the mean motion $n=2\pi/87.96935$ day$^{-1}$ and the sidereal frequency $\Omega_o=2\pi/58.64623$ day$^{-1}$, with $\Omega_o=1.5 \,n$.  Mercury's rotational state is also characterized by a Cassini state whereby the orientations of the orbit normal ($\mitbf{\hat{e}_3^I}$) and of the mantle symmetry axis ($\mitbf{\hat{e}_3^p}$) are both coplanar with, and precess about, the normal to the Laplace plane ($\mitbf{\hat{e}_3^L}$). The orientation of the Laplace plane varies on long timescales, but it can be taken as invariable in inertial space for our present purpose.  The Cassini state of Mercury is illustrated in Figure \ref{fig:cassini}.  The angle between $\mitbf{\hat{e}_3^L}$ and $\mitbf{\hat{e}_3^I}$ is the orbital inclination $I=8.5330^\circ$ \cite[][]{baland17}, the angle between $\mitbf{\hat{e}_3^I}$ and $\mitbf{\hat{e}_3^p}$ is the obliquity $\varepsilon_m$ and the angle between  $\mitbf{\hat{e}_3^L}$ and $\mitbf{\hat{e}_3^p}$ is $\theta_p=I+\varepsilon_m$.  The precession of $\mitbf{\hat{e}_3^I}$ and $\mitbf{\hat{e}_3^p}$ about the Laplace pole is retrograde with frequency $\Omega_p = 2\pi/325,513$ yr$^{-1}$ \cite[][]{baland17}.

The mantle and crust are welded together and form a single rotating region which we refer to as the `mantle' in the context of our rotational model.  The rotation and symmetry axes of the mantle are expected to remain in close alignment, but they do not coincide exactly.  We define the rotation rate vector of the mantle by $\boldsymbol{\Omega}$, and its misalignment from $\mitbf{\hat{e}_3^p}$ by an angle $\theta_m$.  Note that $\theta_m\ll \varepsilon_m$ and it is often the spin axis of Mercury which is used to define the obliquity $\varepsilon_m$ \cite[e.g.][]{baland17}.  If Mercury were an entirely rigid planet, $\mitbf{\hat{e}_3^p}$ and $\boldsymbol{\Omega}$ would characterize the symmetry and rotation axes of the whole of Mercury, not just its mantle, and the angles $I$, $\varepsilon_m$ and $\theta_m$ would completely describe the Cassini state.  The presence of a fluid outer core and solid inner core require three additional orientation vectors and angles.  The symmetry axis of the inner core is defined by unit vector $\mitbf{\hat{e}_3^s}$ and its misalignment from $\mitbf{\hat{e}_3^p}$ by an angle $\theta_n$.  The rotation vectors of the fluid core and inner core are defined as $\boldsymbol{\Omega_f}$ and $\boldsymbol{\Omega_s}$, respectively, and their misalignment from the rotation vector of the mantle $\boldsymbol{\Omega}$ are defined by angles $\theta_f$ and $\theta_s$ (see Figure \ref{fig:angles}a). The rotation and symmetry axes of the inner core remain in close alignment, so $\theta_n \approx \theta_s$.  To be formal in our definition of the different angles of misalignment, for $I$ defined positive as depicted on Figure \ref{fig:angles}a, all angles are defined positive in the clockwise direction.

At equilibrium in the Cassini state, the three orientation vectors ($\mitbf{\hat{e}_3^I}$, $\mitbf{\hat{e}_3^p}$, $\mitbf{\hat{e}_3^s}$) and three rotation vectors ($\boldsymbol{\Omega}$, $\boldsymbol{\Omega_f}$, $\boldsymbol{\Omega_s}$) are forced to precess about $\mitbf{\hat{e}_3^L}$ at the same frequency.  If we neglect dissipation, all vectors lie on the same plane, which we refer to as the Cassini plane.  Viewed in inertial space, the Cassini plane is rotating in a retrograde direction at frequency $\Omega_p$.   Viewed in the frame attached to the mantle rotating at sidereal frequency $\Omega_o$, the Cassini plane is rotating in a retrograde direction at frequency $\omega \Omega_o$ (see Figure \ref{fig:angles}b), where $\omega$, expressed in cycles per Mercury day, is equal to 

\begin{equation}
\omega = -1 - \delta \omega \cos (\theta_p) \, .
\label{eq:omega1}
\end{equation}
The factor $\delta \omega = \Omega_p/\Omega_o = 4.933 \times 10^{-7}$ is the Poincar\'e number, expressing the ratio of the forced precession to sidereal rotation frequencies.   The invariance of the Laplace plane normal as seen in the mantle frame is expressed as 

\begin{equation}
 \frac{d}{dt}  \mitbf{\hat{e}_3^L}  + \boldsymbol{\Omega} \times  \mitbf{\hat{e}_3^L}  = {\bf 0} \, ,\label{eq:de3Ldt}
\end{equation}
or equivalently, by Equation (19e) of \cite{stys18},

\begin{equation}
\omega \sin (\theta_p) + \sin (\theta_m + \theta_p) = 0 \, . \label{eq:kinL1}
\end{equation}
This expresses a formal connection between $\theta_p$ and $\theta_m$ which is independent of the interior structure of Mercury.  Using Equation (\ref{eq:omega1}) and $\cos({\theta_m}) \rightarrow 1$, this connection can be rewritten as  

\begin{equation}
\sin (\theta_m) = \delta \omega \, \sin (\theta_p) \, .\label{eq:sinm}
\end{equation}
and thus the relative amplitudes of $\theta_m$ and $\theta_p$ depend of the Poincar\'e number $\delta \omega$.  

To investigate Mercury's response to the gravitational torque from the Sun, we take advantage of the framework developed in \citet{mathews91a} to model the forced nutations of Earth \cite[see also][]{mathews02,dehant15}.  This model takes into account the pressure torque (also referred to as the inertial torque) that results when the spin axis of the fluid core is misaligned from the symmetry axes of the elliptical surfaces of the CMB and ICB.  It also includes the gravitational torque exerted on the inner core when it is misaligned with the mantle.  Electromagnetic and viscous torques at both the CMB and ICB have been incorporated into the framework \cite[e.g][]{buffett92,buffett02,mathews05,deleplace06}.  The framework was adapted to model the Cassini state of the Moon in \cite{dumberry16} and further developed in \cite{stys18} and \citet{organowski20}.   We adapt it here to capture the Cassini state of Mercury. 

Because the forced precession period is much longer than the rotation and orbital periods of Mercury, the gravitational solar torque that is relevant to the Cassini state is the mean torque averaged over one orbit.  This mean torque is perpendicular to the Cassini plane, pointing in the same direction as the vector connecting the Sun to the descending node of Mercury's orbit in Figure 1.  Hence, viewed from the mantle frame, the orientation of this mean torque is periodic, rotating at frequency $\omega \Omega_o$.  Setting the equatorial directions $\mitbf{\hat{e}_1^p}$ and $\mitbf{\hat{e}_2^p}$ to correspond to the real and imaginary axes of the complex plane, respectively, we can write the equatorial components of this periodic applied torque in a compact form as 

\begin{equation}
{\Gamma}_1(t) + i {\Gamma}_2(t) = - i  \, \tilde{\Gamma}(\omega) \,  \exp[{i \omega \Omega_o t}] \, , \label{eq:gammaphi}
\end{equation}
where $\tilde{\Gamma}(\omega)$ represents the amplitude of the torque at frequency $\omega \Omega_o$.  In response to this torque, the axes defining all angles ($\theta_p$, $\varepsilon_m$, $\theta_m$, $\theta_f$, $\theta_s$, $\theta_n$) as viewed in the mantle frame are also rotating at frequency $\omega \Omega_o$ (see Figure \ref{fig:angles}).  The longitudinal direction of each of these angles at a specific time $t$ can then also be written in the equatorial complex plane and is proportional to $\exp[{i \omega \Omega_o t}]$.  For instance, the two equatorial time-dependent components $\theta_{m1}$ and $\theta_{m2}$ of the angle $\theta_m$, as seen in the mantle frame, can be written as

\begin{subequations}
\begin{equation}
\theta_{m1}(t) + i \theta_{m2}(t) =  \tilde{m} \,  \exp[{i \omega \Omega_o t}] \, , 
\end{equation} 
where 

\begin{equation}
\tilde{m} \equiv \tilde{m}(\omega) = Re[ \tilde{m}] + i Im[\tilde{m}] \, , 
\end{equation} 
\end{subequations}
is the amplitude at frequency $\omega \Omega_o$.  Equivalent definitions apply for all other angles, with the connection as follows:

\begin{equation}
\theta_m \Leftrightarrow \tilde{m} \, , \hspace*{0.5cm}
\theta_f \Leftrightarrow \tilde{m}_f \, , \hspace*{0.5cm}
\theta_s \Leftrightarrow \tilde{m}_s \, , \hspace*{0.5cm}
\theta_n \Leftrightarrow \tilde{n}_s \, , \hspace*{0.5cm}
\theta_p \Leftrightarrow \tilde{p} \, , \hspace*{0.5cm}
\varepsilon_m  \Leftrightarrow \tilde{\varepsilon}_m \, . 
\end{equation}
The notation $\tilde{m}$, $\tilde{m}_f$, $\tilde{m}_s$, $\tilde{n}_s$ follows that introduced in the original model of \cite{mathews91a}.  Note that all tilded amplitudes are complex: their imaginary part reflects the out-of-phase response to the applied torque as a result of dissipation, for instance from viscous or EM coupling at the boundaries of the fluid core.  In the absence of dissipation, all tilded variables are purely real.  We concentrate our analysis in this work on the real part of the solutions, which corresponds to the mutual alignment of these five rotation angles in the Cassini plane.  As such, $\tilde{\varepsilon}_m$ corresponds to the observed obliquity of the mantle symmetry axis. It is thus equivalent to $\varepsilon_m$, though we keep the tilde notation in the presentation of our results  to emphasize that it represents the real part of the solution from our system.  Furthermore, since $\tilde{m}\ll \tilde{\varepsilon}_m$, we often refer to $\tilde{\varepsilon}_m$ as the orientation of spin axis of the mantle, since the Cassini state of Mercury is more customarily described in terms of the latter in the literature.

The model of \cite{mathews91a} is developed under the assumption of small angles as appropriate for the nutations on Earth.  The details on how the equations of the model are derived can found in \cite{mathews91a} and in \cite{dumberry16}.  Three equations describe, respectively,  the time rate of change of the angular momenta of the whole of Mercury, the fluid core, and the inner core in the reference frame of the rotating mantle.   These three equations are

\begin{subequations}
\begin{equation}
 (\omega - e)\tilde{m} + (1 +\omega) \Bigg[ \frac{\bar{A}_f}{\bar{A}} \tilde{m}_f + \frac{\bar{A}_s}{\bar{A}}  \tilde{m}_s + \alpha_3 e_s \frac{\bar{A}_s}{\bar{A}} \tilde{n}_s  \Bigg] = \frac{1}{i \Omega_o^2 \bar{A}} \Big(\tilde{\Gamma}_{sun}  \Big) \, , \label{eq:am1}
 \end{equation}
 
 \begin{equation}
 \omega \tilde{m} + \left( 1 + \omega  + e_f  \right) \tilde{m}_f -  \omega \alpha_1 e_s \frac{\bar{A}_s}{\bar{A}_f} \tilde{n}_s  =   \frac{1}{i \Omega_o^2 \bar{A}_f} \Big(- \tilde{\Gamma}_{cmb}  - \tilde{\Gamma}_{icb} \Big) \, , \label{eq:af1}
 \end{equation}
 
 \begin{equation}
( \omega - \alpha_3 e_s) \tilde{m} + \alpha_1 e_s \tilde{m}_f + \left(1+\omega\right) \tilde{m}_s +  \left(1+\omega - \alpha_2 \right) e_s \tilde{n}_s = \frac{1}{i \Omega_o^2 \bar{A}_s} \Big( \tilde{\Gamma}_{sun}^s  +  \tilde{\Gamma}_{icb} \Big) \, , \label{eq:as1}  
 \end{equation}
and a fourth equation consists of a kinematic relation that expresses the change in the orientation of the inner core figure as a result of its own rotation,
 
\begin{equation}
\tilde{m}_s  + \omega  \tilde{n}_s  = 0 \, . \label{eq:msns}
\end{equation}
\label{eq:sys4}
\end{subequations}

In these equations, the parameters $\alpha_1$, $\alpha_2$ and $\alpha_3$ involve the density contrast at the ICB and are given by 

\begin{subequations}
\begin{equation}
\alpha_1 = \frac{\rho_f}{\rho_s} \, , \hspace*{0.5cm} \alpha_3  = 1- \alpha_1  \, , \hspace*{0.5cm}
\alpha_2 = \alpha_1 - \alpha_3 \alpha_g  \, , 
\end{equation}
where the parameter $\alpha_g$ is a measure of the ratio of the gravitational to inertial torque applied on the inner core,
 
\begin{equation}
\alpha_g  = \frac{8\pi G}{5\Omega_o^2} \left[ \rho_c (\epsilon_r - \epsilon_m) + \rho_m (\epsilon_m - \epsilon_f) + \rho_f \epsilon_f \right] \, ,
\end{equation} 
\label{eq:alpha132g}
\end{subequations}
where $G$ is the gravitational constant.  

$\tilde{\Gamma}_{sun}$ is the amplitude of the gravitational torque by the Sun on the whole of Mercury.  For a small mantle obliquity $\tilde{\varepsilon}_m$ and a small inner core tilt $\tilde{n}_s$, it is given by

\begin{equation}
\tilde{\Gamma}_{sun}  = -i \Omega_o^2 \bar{A} \left( \phi_m  \tilde{\varepsilon}_m  + \frac{\bar{A}_s}{\bar{A}} \alpha_3 \phi_s \tilde{n}_s  \right) \, ,\label{eq:tqsun}
\end{equation}
where

\begin{subequations}
\begin{align}
\phi_m & = \frac{3}{2} \frac{n^2}{\Omega_o^2} \left[ G_{210} \, e + \frac{1}{2} G_{201} \, \gamma  \right] \, , \label{eq:phim}\\
\phi_s & = \frac{3}{2} \frac{n^2}{\Omega_o^2} \left[ G_{210} \,  e_s   + \frac{1}{2} G_{201}\, \gamma_s   \right] \, ,\label{eq:phis}
\end{align}
\label{eq:phims}
 \end{subequations}
and where  $G_{210}$ and $G_{201}$ are functions of the orbital eccentricity $e_c$,

\begin{subequations}
\begin{align}
G_{210} & = \frac{1}{(1-e_{c}^2)^{3/2}} \, ,\label{eq:G210} \\
G_{201}  & = \frac{7}{2} e_c  - \frac{123}{16} e_c^3 + \frac{489}{128} e_c^5 \, .\label{eq:G201}
\end{align}
 \end{subequations}
The gravitational torque by the Sun acting on the inner core alone, $\tilde{\Gamma}_{sun}^s$, is 

\begin{equation}
\tilde{\Gamma}_{sun}^s  =  -i \Omega_o^2 \bar{A}_s \alpha_3 \phi_s (\tilde{\varepsilon}_m + \tilde{n}_s)  \, .
\label{eq:tqsuns}
\end{equation}
$\tilde{\Gamma}_{cmb}$ and $\tilde{\Gamma}_{icb}$ are the torques from tangential stresses by the fluid core on the mantle at the CMB and on the inner core at the ICB, respectively.   These torques can be parameterized in terms of dimensionless complex coupling constants $K_{icb}$ and $K_{cmb}$ and the differential angular velocities at each boundary \cite[e.g][]{buffett92,buffett02},

\begin{subequations}
\begin{equation}
    \tilde{\Gamma}_{icb} = i \Omega_o^2 \bar{A}_s K_{icb} (\tilde{m}_f - \tilde{m}_s) \, ,\label{eq:tqicb}
\end{equation}
\begin{equation}
    \tilde{\Gamma}_{cmb} = i \Omega_o^2 \bar{A}_f K_{cmb} \, \tilde{m}_f \, .\label{eq:tqcmb}
\end{equation}
\end{subequations}
Specific expressions for $K_{icb}$ and $K_{cmb}$ are delayed to sections 4 and 5 when we consider the effects of viscous and EM coupling, respectively.

A fifth equation is required to connect this interior model to the obliquity of the mantle, and this is provided by Equation (\ref{eq:kinL1}).   For small angles $\theta_m$ and $\theta_p$, this gives \cite[e.g.][]{mathews91a,dumberry16,baland19}
   
\begin{equation}
\tilde{m} + (1+\omega) \tilde{p} = 0 \, . \label{eq:kinmp}
 \end{equation}
For Mercury, it is more convenient to connect the internal model with $\tilde{\varepsilon}_m$ instead of $\tilde{p}$.  This is because $\theta_p\approx8.567^\circ$ whereas $\tilde{\varepsilon}_m \approx 2$ arcmin and thus the latter obeys more strictly the condition of small angles assumed in our framework.  Furthermore, the external torques acting on the whole planet (Equation \ref{eq:tqsun}) and inner core (Equation \ref{eq:tqsuns}) depend linearly on $\tilde{\varepsilon}_m$.   Written in terms of $\tilde{\varepsilon}_m$, and with the approximation of $\tilde{\varepsilon}_m \ll 1$ and $\tilde{m} \ll 1$, Equation (\ref{eq:kinL1}) becomes 

\begin{equation}
\tilde{m} + (1+\omega) \tilde{\varepsilon}_m  = - (1+\omega)  \tan I  \, .
 \label{eq:kinI}
 \end{equation}
Likewise, the frequency $\omega$ from Equation (\ref{eq:omega1}) can be written simply in terms of $I$,

\begin{equation}
\omega = -1 - \delta \omega \cos I \, .
\label{eq:omega}
\end{equation}

The set of four Equations (\ref{eq:sys4}) with the addition of Equation (\ref{eq:kinI}) form a linear system of equations for the five rotational variables $\tilde{m}$, $\tilde{m}_f$,  $\tilde{m}_s$, $\tilde{n}_s$ and $\tilde{\varepsilon}_m$.  It captures the response of Mercury, in the frequency domain, when subject to a periodic solar torque applied at frequency $\omega$.  The system can be written in a matrix form as

\begin{subequations}
\begin{equation}
\boldsymbol{\mathsf{M}} \cdot {\bf x} \, = {\bf y} \, ,
\label{eq:mat}
\end{equation}
where the solution (${\bf x}$) and forcing (${\bf y}$) vectors are 

\begin{align}
{\bf x}^T &= \left[ \tilde{m}, \tilde{m}_f, \tilde{m}_s, \tilde{n}_s, \tilde{\varepsilon}_m \right] \, ,\\ 
{\bf y}^T &= \left[ 0, 0, 0, 0, -(1+\omega) \tan I\right] \, ,
\end{align}
and the elements of matrix $\boldsymbol{\mathsf{M}}$ are 
 
\begin{equation}
\boldsymbol{\mathsf{M}} = \begin{bmatrix}  \omega - e & (1 + \omega)  \frac{\bar{A}_f}{\bar{A}}  &  (1 + \omega) \frac{\bar{A}_s}{\bar{A}} &\frac{\bar{A}_s}{\bar{A}} \alpha_3 \big( (1 + \omega) e_s + \phi_s\big)   & \phi_m \\
\omega & 1 + \omega  + e_f + K_{cmb} +  \frac{\bar{A}_s}{\bar{A}_f} K_{icb} &  - \frac{\bar{A}_s}{\bar{A}_f} K_{icb} &  -\omega e_s \alpha_1 \frac{\bar{A}_s}{\bar{A}_f}  & 0 \\
\omega - \alpha_3 e_s  &  \alpha_1 e_s  - K_{icb}  & 1 + \omega + K_{icb} & (1 + \omega -\alpha_2) e_s  + \alpha_3 \phi_s & \alpha_3 \phi_s \\
0 & 0 & 1 & \omega & 0 \\
1 & 0 & 0 & 0 & (1 + \omega)
\end{bmatrix} \, . 
\end{equation}
\end{subequations} 

Solutions of the homogeneous system (i.e. ${\bf y=0}$) represent free modes of precession.  Three modes have periods which, when seen in inertial space, are typically in the range of a few hundred to a few thousand years.  The first is the free axial precession of Mercury maintained by the solar torque acting on its elliptical figure \cite[e.g.][]{peale05}.  The second is the free core nutation (FCN), which is the free precession of the spin axis of the fluid core about the symmetry axis of the CMB \cite[e.g.][]{mathews91a}.  The third is the free inner core nutation (FICN), a free mode of rotation similar to the FCN but associated with the inner core \cite[e.g.][]{mathews91a}.  

A few remarks on our model are important to point out before we proceed further.   First, although we have retained the triaxial shape of Mercury in the expression of the solar torque, we treat its angular momentum response as if it were an axially symmetric body.  This is convenient as the two equatorial angular momentum equations for each region can be combined into a single equation.  To first order, the frequency of the free precession of Mercury is not largely altered by triaxiality \cite[e.g.][]{peale05}. \citet{baland19} showed that the frequencies of the FCN and FICN for a triaxial planetary body may be slightly different than those for an axially symmetric body, but not by large factor.  As the response of Mercury to the solar torque is largely determined by the resonant amplification due to the presence of these three modes, our model should capture correctly the first order Cassini state of Mercury.  Considering the triaxial shape of Mercury may alter the numerical results, but not our general conclusions.

Second, our modelling approach is different than in the studies of \cite{peale14} and \cite{peale16}.  In these two studies, dynamical models of Mercury's Cassini state are developed and must then be integrated in time.  The equilibrium Cassini state is the quasi-steady state that remains after transient effects associated with the initial conditions have decayed away.   An advantage of these models compared to ours is that the complete triaxial dynamics of Mercury, including its longitudinal librations, are retained.   However, the numerical integration can be lengthy if dissipation is weak, which restricts the number of possible interior models of Mercury that can be tested.  In contrast, our model is a simple linear system in the frequency domain, focused on one specific frequency: the forced precession associated with the Cassini state.  Solutions are straightforward to obtain for a given interior model, and this allows us to cover a larger span of the parameter space.  One drawback, however, is that our model does not capture time-dependent variations at any other frequencies, including the precession of the pericenter of Mercury's orbit about the Sun.  
  
 \subsection{Analytical solutions and limiting cases}
 
\subsubsection{The Cassini state of a single-body, rigid Mercury}
 
For a rigid planet with no fluid and solid cores, our system of equations reduces to  Equations (\ref{eq:am1}) and (\ref{eq:kinI}),
 
\begin{subequations}
\begin{align} 
& (\omega - e)\tilde{m} + \phi_m \, \tilde{\varepsilon}_m = 0 \, , \label{eq:angmomrigid}\\
& \tilde{m} + (1+\omega) \tilde{\varepsilon}_m  = - (1+\omega)  \tan I  \, . \label{eq:tildem1}
\end{align}
\label{eq:sysrigid}
\end{subequations}
Using Equation (\ref{eq:omega}), $\delta \omega \ll1$, and the approximation $\bar{A}(1+e + \delta \omega \cos I) = C + \bar{A} \delta \omega \cos I \approx C$, these can be written as

\begin{subequations}
\begin{align} 
& C \tilde{m} = \bar{A} \phi_m \, \tilde{\varepsilon}_m  \, , \label{eq:angmomrigid2}\\
& \tilde{m} = \delta \omega \big(\sin I + \cos I \, \tilde{\varepsilon}_m \big)   \,. \label{eq:tildem1}
\end{align}
\label{eq:sysrigid2}
\end{subequations}

Equation (\ref{eq:tildem1}) gives a direct relationship between $\tilde{m}$ and $\tilde{\varepsilon}_m$.  For $I=8.5330^\circ$, $\delta \omega = 4.9327 \times 10^{-7}$ and taking $\tilde{\varepsilon}_m=2.04$ arcmin, this gives $\tilde{m}=2.52 \times 10^{-4}$ arcmin, much smaller than $\tilde{\varepsilon}_m$: the offset of the rotation axis of the mantle with respect to its symmetry axis is very small.   Substituting Equation (\ref{eq:tildem1}) in Equation (\ref{eq:angmomrigid2}) gives

\begin{equation}
C \Omega_p \big(\sin I + \cos I \, \tilde{\varepsilon}_m \big) = \bar{A} \Omega_o \phi_m \tilde{\varepsilon}_m \, , \label{eq:amrigid}
\end{equation}
and isolating for $\tilde{\varepsilon}_m$,

\begin{equation}
\tilde{\varepsilon}_m = \frac{{C} \Omega_p \sin I}{-{C} \Omega_p \cos I +  \bar{A} \Omega_o \phi_m } \, .
\label{eq:predepsm1}
\end{equation}
Upon using Equations (\ref{eq:egammaJ2C22}), (\ref{eq:phim}), and $\Omega_o = \frac{3}{2} n$, we can write 

\begin{equation}
\tilde{\varepsilon}_m = \frac{{C} \Omega_p \sin I}{-{C} \Omega_p \cos I + n MR^2 \left( G_{210} J_2 + 2  G_{201}C_{22} \right)} \, .
\label{eq:predepsm}
\end{equation}
This is the standard prediction for the obliquity of a rigid Mercury occupying Cassini state 1 \cite[see for instance Equation (1) of][where their definition of $\dot{\Omega}$ is equal to $-\Omega_p$]{baland17}.  Hence, in the absence of a fluid core and inner core, our system retrieves the Cassini state of Mercury correctly.  Equation (\ref{eq:predepsm}) can be manipulated to solve instead for the normalized moment of inertia $\hat{C}$, 

\begin{equation}
\hat{C} = \frac{C}{MR^2} = \frac{n}{\Omega_p} \frac{ G_{210} J_2 + 2  G_{201} C_{22}}{  \cos I + \sin I/\tilde{\varepsilon}_m }\, . \label{eq:ctilde}
\end{equation}
which is equivalent to Equation (89) of \cite{vanhoolst15}.  It is based on the latter equation that a measurement of the obliquity gives a constraint on $\hat{C}$.

Two free modes of precession are found by setting ${\bf y=0}$ in Equation (\ref{eq:sysrigid}).  One mode corresponds to the Eulerian wobble, or Chandler wobble, and represents the prograde precession of the rotation axis about the symmetry axis.  The second mode is the free retrograde axial precession of Mercury.  As seen in the inertial frame, its frequency is given by 

\begin{equation}
\omega_{fp} = n \frac{MR^2}{C} \Big( G_{210} J_2 + 2  G_{201} C_{22} \Big) \, ,\label{eq:fp}
\end{equation}
which is equivalent to the prediction by \cite{peale05} when neglecting its small elliptical component.  Note that in \cite{peale05} it was assumed that only the mantle was involved in the solid-body precession and hence $C$ was replaced by $C_m$.  Using $C=0.346 \cdot M R^2$ \cite[][]{margot12} and the numerical values for $n$, $J_2$, $C_{22}$ and $e_c$ given in Table 1, we obtain a free precession period of $T_{fp}=2\pi/\omega_{fp} = 1298$ yr.  If we use $C_m$ instead of $C$ in Equation (\ref{eq:fp}), and take $C_m=0.431 \cdot C = 0.431 \cdot 0.346 \cdot MR^2$ \cite[][]{margot12}, we obtain $T_{fp}=2\pi/\omega_{fp} = 560$ yr.  These estimates are similar to those obtained by \cite{peale05}.  Because the CMB is elliptical, the pressure torque exerted on the fluid core by the mantle leads to an entrainment of the fluid core, the degree of which depends on the amplitude of the pole-to-equator CMB flattening.  The true free precession period lies somewhere between 560 and 1298 yr.  Regardless of its exact value, the free precession period is much shorter than the forcing period of 325 kyr.  Using Equation (\ref{eq:fp}), Equation (\ref{eq:predepsm}) can be written as \cite[e.g.][]{baland17}

\begin{equation}
\tilde{\varepsilon}_m = \frac{ \Omega_p \sin I}{- \Omega_p \cos I +  \omega_{fp} } \, .
\label{eq:predepsmfp}
\end{equation}
The obliquity of Mercury is thus determined by how the forcing frequency $\Omega_p$ compares with the free precession frequency $\omega_{fp}$.  Because $\omega_{fp} > \Omega_p$, Mercury occupies Cassini state 1 \cite[][]{peale74}.  Furthermore, Equation (\ref{eq:predepsmfp}) shows that a large obliquity can be generated by resonant amplification if $\Omega_p\approx\omega_{fp}$.  Since $\omega_{fp} \gg \Omega_p$, resonant amplification is minimal and the resulting obliquity, $\tilde{\varepsilon}_m\approx 2$ arcmin, is much smaller than the inclination angle $I\approx 8.5^\circ$.  

\subsubsection{The misalignment of the fluid and solid cores}

With $\omega=-1 -\delta\omega \cos I $ and $\delta \omega \ll1$, Equation (\ref{eq:msns}) gives $\tilde{n}_s\approx \tilde{m}_s$; as for the mantle, the rotation and symmetry axes of the inner core remain closely aligned in the Cassini state.  The relationship between $\tilde{m}$ and $\tilde{\varepsilon}_m$ of Equation (\ref{eq:tildem1}) is independent of the interior structure, so it remains unchanged when a fluid and a solid cores are present. Substituting it in Equation (\ref{eq:am1}), and setting $\tilde{n}_s=\tilde{m}_s$, the angular momentum equation of the whole planet becomes

\begin{equation}
C \Omega_p \big(\sin I + \cos I \, \tilde{\varepsilon}_m \big)  + (\bar{A}_f \cos I \, \Omega_p ) \tilde{m}_f + \bar{A}_s( \cos I \, \Omega_p-  \Omega_o\alpha_3 \phi_s)\tilde{n}_s = \bar{A} \Omega_o  \phi_m \tilde{\varepsilon}_m \, . \label{eq:angmomplanet}
\end{equation}
This latter equation shows how the misaligned inner core and fluid core can lead to a modification of the mantle obliquity $\tilde{\varepsilon}_m$.   Approximate analytical solutions of $\tilde{n}_s$ and $\tilde{m}_f$ are given by 

\begin{subequations}
\begin{align}
\tilde{n}_s & \approx \frac{\Omega_p}{\kappa \lambda_s}  \left(1 + \frac{\Omega_o ( K_{icb}-\alpha_1 e_s)}{\lambda_f} \right) \big(\sin I + \cos I \, \tilde{\varepsilon}_m \big)  - \frac{\Omega_o \alpha_3\phi_s}{\kappa \lambda_s} \tilde{\varepsilon}_m \, ,\\
\tilde{m}_f & \approx \frac{\Omega_p}{\lambda_f}  \big(\sin I + \cos I \, \tilde{\varepsilon}_m \big) + \frac{\Omega_o}{\lambda_f} \frac{\bar{A}_s}{\bar{A}_f} \big( K_{icb} - \alpha_1 e_s \big) \tilde{n}_s \, , \label{eq:approxmf}
\end{align}
\end{subequations}
where
 
\begin{subequations}
\begin{align}
\kappa & =1 -   \frac{\bar{A}_s}{\bar{A}_f} \frac{\Omega_o^2 \big( K_{icb} - \alpha_1 e_s \big)^2}{\lambda_s \, \lambda_f} \, ,\\
\lambda_f & = \bar{\sigma}_f -\Omega_p \cos I  \label{eq:lambdaf} \, ,\\ 
\lambda_s & = \bar{\sigma}_s -\Omega_p \cos I  \label{eq:lambdas} \, ,
\end{align}
and where we have introduced the frequencies

\begin{align}
\bar{\sigma}_f & = \Omega_o\left(e_f + K_{cmb} +\frac{\bar{A}_s}{\bar{A}_f} K_{icb} \right) \label{eq:sigmaf}\, , \\ 
\bar{\sigma}_s & =  \Omega_o\Big(e_s \alpha_3 \alpha_g - e_s \alpha_1 + \alpha_3 \phi_s + K_{icb}  \Big)   \label{eq:sigmas} \, .
\end{align}
\end{subequations}
These solutions are good approximations for all the results that we present in section 3.  For an observed mantle obliquity $\tilde{\varepsilon}_m$ and for a chosen set of interior model parameters, they provide useful predictions of $\tilde{n}_s$ and $\tilde{m}_f$. 

In the limit of a very strong coupling between the fluid core, solid core and mantle, $\bar{\sigma}_s \gg \Omega_p$ and  $\bar{\sigma}_f \gg \Omega_p$, so that $\tilde{n}_s \rightarrow  0$, $\tilde{m}_f  \rightarrow  0$ and Equation (\ref{eq:angmomplanet}) reverts back to Equation (\ref{eq:amrigid}) for a rigid planet.  In the opposite limit of no coupling between  the fluid core, solid core and mantle (i.e. for spherical internal boundaries, $e_f=e_s=\gamma_s=0$ and no viscous or EM coupling, $K_{cmb}=K_{icb}=0$), then

\begin{equation}
\phi_s = 0 \,, \hspace*{0.5cm} \kappa=1\, , \hspace*{0.5cm} \lambda_f=\lambda_s= - \Omega_p\cos I \, , \hspace*{0.5cm}
\tilde{m}_f = \tilde{n}_s = - (\tan I +  \tilde{\varepsilon}_m ) \, .
\end{equation}
Inserting these in Equation (\ref{eq:angmomplanet}), and with the moment of inertia of the mantle equal to $C_m= C -\bar{A}_f - \bar{A}_s$, we obtain 

\begin{equation}
C_m\, \Omega_p \big(\sin I + \cos I \, \tilde{\varepsilon}_m \big)  = \bar{A} \Omega_o \phi_m \tilde{\varepsilon}_m \, . \label{eq:angmommantle}
\end{equation}
which describes, as expected, a forced precession of the mantle alone. If this was the case for Mercury, taking $C_m/C=0.431$, the obliquity should be $\tilde{\varepsilon}_m\approx0.88$ arcmin, substantially smaller than the observed obliquity of  $\tilde{\varepsilon}_m\approx2$ arcmin.  

If $\bar{\sigma}_f \approx \Omega_p$ (and thus $\lambda_f\rightarrow0$) and/or $\bar{\sigma}_s \approx \Omega_p$ (and thus $\lambda_s\rightarrow0$) resonant amplification leads to large amplitudes for $\tilde{m}_f$, $\tilde{n}_s$ and the mantle obliquity $\tilde{\varepsilon}_m$. The frequencies $\bar{\sigma}_f$ and $\bar{\sigma}_s$ are closely related to the FCN and FICN frequencies $\omega_{fcn}$ and $\omega_{ficn}$, respectively.  Hence, just as a large mantle obliquity can result from resonant amplification when the forcing frequency approaches the free precession frequency, a large mantle obliquity can likewise result from resonant amplification when the forcing frequency approaches the FCN or FICN frequencies.   These frequencies depend on the interior density structure and are not known.  However, we will show that for reasonable interior models of Mercury, the FCN and FICN periods are in the range of a few hundred yr.  This is sufficiently far from the forcing period (325 kyr) that we do not expect an important amplification effect.  Furthermore, since $\omega_{fcn}, \omega_{ficn} \gg \Omega_p$, then $\bar{\sigma}_f \gg \Omega_p$ and $\bar{\sigma}_s \gg \Omega_p$, and we are in the strong coupling limit.  The mantle obliquity should be close to that expected for a rigid planet, as observations suggest.   Therefore, we expect  that $\tilde{m}_f$ and $\tilde{n}_s$ should be of the order of $\tilde{\varepsilon}_m$ or smaller.  This further justifies the assumption of small angles that we have adopted.

\section{Results}

\subsection{Geodetic constraints and interior density structure}

All our interior models are constrained to match the mass $M$ of Mercury and specific choices of $\hat{C}=C/MR^2$ and $C_m/C$.  The choice of $\hat{C}$ is determined from Equation (\ref{eq:ctilde}).  
For the parameters listed in Table \ref{tab:para}, and an observed obliquity of $\varepsilon_m=2.04$ arcmin \cite[][]{margot12}, this gives $\hat{C}=C/MR^2 = 0.3455$ and all our interior models are consistent with this choice.  Obviously, this reflects a Cassini state equilibrium in which the fluid core and inner core are perfectly aligned with the mantle, which is not strictly correct.   Hence, we make an error in estimating  $\hat{C}$ from Equation (\ref{eq:ctilde}), or conversely in predicting $\varepsilon_m$ based on a given choice for $\hat{C}$.  Part of the objective of our study is to estimate how large this error is.  The ratio $C_m/C$ is obtained from the amplitude of the 88-day longitudinal mantle libration $\phi_o$, which is given by  

\begin{equation}
\phi_o = 6 \cdot f(e_c) C_{22} \frac{MR^2}{C}\frac{C}{C_m} \frac{1}{1+\zeta}\, , \label{eq:phio}
\end{equation}
where 

\begin{equation}
f(e_c) = 1 - 11 e_c^2 + \frac{959}{48} e_c^4 \, ,
\end{equation} 
and where $\zeta$ is a correction that takes into account the entrainment of the inner core in the libration \cite[][]{vanhoolst12,dumberry13,dumberry15}; this correction is small and, to simplify, we neglect it here.  Taking the observed libration amplitude to be 38.5 arcsec \cite[][]{margot12}, $\hat{C}=C/MR^2 = 0.3455$ and $C_{22}$ and $e_c$ from Table \ref{tab:para}, this corresponds to a ratio $C_m/C = 0.4269$, or equivalently $\hat{C}_m=C_m/MR^2 = 0.1475$.
   
For all results presented in our study, the crustal density is set at $\rho_c=2974$ kg m$^{-3}$ \cite[][]{sori18}. Our standard choice for the crustal thickness is $h=26$ km \cite[][]{sori18}, although in section 3.2 we also present some results with other choices of  $h$.  We have considered two possible prescriptions connected to the density of the inner core.  First, for all the results presented in sections 3.2, 3.3 and 3.4, we have used a fixed inner core density of $\rho_s=8800$ kg m$^{-3}$ approximately that obtained in \cite{dumberry15} under the assumption of a pure Fe composition in face-centered cubic phase.  This captures an end-member scenario where the core composition is an Fe-S alloy; at Mercury's core conditions, crystallization of Fe is relatively free of S on the Fe-rich side of the eutectic \cite[][]{li01}.  If the core composition is instead an Fe-Si alloy, approximately equal partitioning of Si between the liquid and solid phase \cite[e.g.][]{schaefer17} implies a weak chemical contrast at the ICB.   The density jump across the ICB is expected to be small, although since density increases with depth, the contrast between the mean densities of the fluid and solid cores is larger.  It is these mean densities that enter our Mercury model with uniform density layers.  To capture this other end-member core composition scenario, in section 3.5 we present results where we instead prescribe a fixed density contrast between the fluid and solid core; specifically, we set the numerical value of $\alpha_3$.

\begin{figure}
\begin{center}
\centerline{ \includegraphics[height=5.5cm]{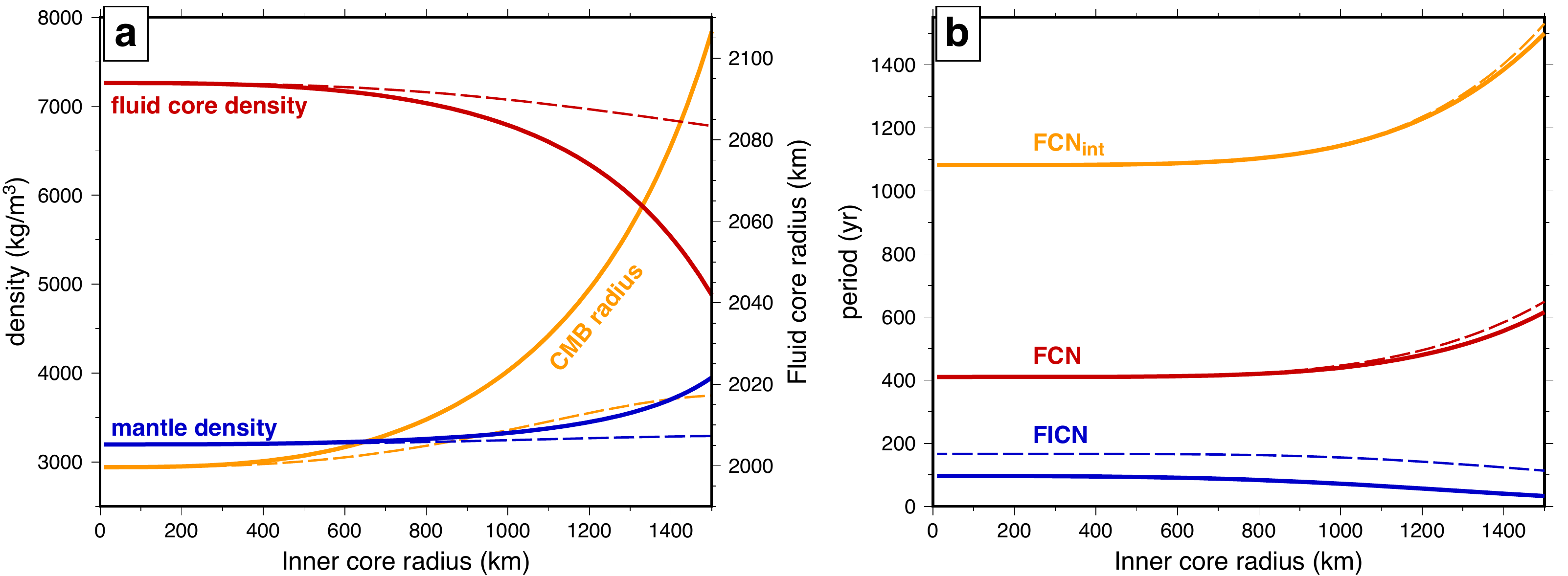} }
\caption{\label{fig:dens} a) Fluid core density (red), mantle density (blue), fluid core radius (orange, right-hand side scale) and b) FICN (blue) and FCN (red) periods as a function of inner core radius.  The FCN period when the external torque is set to zero (FCN$_{\mbox{\small{int}}}$) is shown in orange.  Solid lines correspond to a scenario where the density of the inner core is set to $8800$ kg m$^{-3}$; thin dashed lines correspond to a scenario where the density contrast between the fluid and solid cores is set to $\alpha_3=0.1$.}
\end{center}
\end{figure}

For a given choice of inner core radius $r_s$, the densities of the mantle ($\rho_m$) and fluid core ($\rho_f$) and the radius of the CMB ($r_f$) are determined such that the interior model matches $M$,  $\hat{C}= 0.3455$ and $\hat{C}_m= 0.1475$.  Figure \ref{fig:dens}a shows how $\rho_m$, $\rho_f$ and $r_f$ vary as a function of inner core radius $r_s$ for each of the two inner core density scenarios: a fixed $\rho_s$, or a fixed $\alpha_3$.  When the inner core is small, its presence has a limited influence on the resulting density structure, and we find $\rho_m=3197$ kg m$^{-3}$, $\rho_f=7263$ kg m$^{-3}$ and $r_f=2000$ km in each of the two scenarios.  When $\rho_s$ is fixed to $8800$ kg m$^{-3}$, as the inner core reaches 1500 km in size, $r_f$ increases to above $2100$ km, $\rho_m$ approaches 4000 kg m$^{-3}$ and $\rho_f$ is reduced to below 5000 kg m$^{-3}$.  Figure \ref{fig:dens}a illustrates that when adopting a fixed $\rho_s$, there is a limit in the possible inner core size, as otherwise $\rho_m$ gets unreasonably large and $\rho_f$ gets inappropriately small (as it would require an excessively large concentration of light elements).  When adopting instead a fixed density contrast, with $\alpha_3=0.1$, the changes in $r_f$, $\rho_m$ and $\rho_f$ with inner core radius are more modest, allowing larger possible inner core sizes.  Different assumptions on $\rho_c$ and $h$ would alter the numerical values shown on Figure \ref{fig:dens}a but not their trends with $r_s$.

Figure \ref{fig:dens}b shows how the FCN and FICN periods vary with $r_s$ for each of the two inner core density scenarios and in the absence of viscous and EM coupling (i.e. $K_{cmb}=K_{icb}=0$).  Both of these free modes are retrograde.  The FCN period is close to 400 yr  for a small inner core, increasing to approximately $600$ yr at the largest $r_s$. The FICN period is shorter, close to 100 yr (160 yr) for a small inner core and decreasing to approximately 40 yr (120 yr) at the largest $r_s$ under the fixed $\rho_s$ (fixed $\alpha_3$) scenario.    This confirms that the FCN and FICN periods are both much shorter than the forcing precession period of 325 kyr and sufficiently far away from it that we do not expect large $\tilde{m}_f$ and $\tilde{n}_s$ from resonant amplification. 

The FCN and FICN periods that we have computed include the influence of the external torque. As shown by \citet{baland19}, the external torque allow solid regions to have a free motion in inertial space thereby affecting the free rotational modes. To a good approximation, the FCN and FICN frequencies (as seen in an inertial frame) for $K_{cmb}=K_{icb}=0$ are given by

\begin{subequations}
\begin{align}
\omega_{fcn} & \approx  -\Omega_o \left(\frac{\bar{A}}{\bar{A}_m + \bar{A}_s} \right) \Big(e_f + \phi_m \Big) +\Omega_o \frac{e_f \phi_m}{(e_f + \phi_m)} \, , \label{eq:fcn} \\
\omega_{ficn} & \approx  \Omega_o \left(\frac{\bar{A} + \bar{A}_s }{\bar{A} - \bar{A}_s}\right)  \Big(e_s \alpha_1 - e_s\alpha_3 \alpha_g - \alpha_3 \phi_s \Big)\, .\label{eq:ficn}
\end{align}
\label{eq:fcnficn}
The expression of the FICN frequency involves the inertial torque (term $e_s \alpha_1$) and the gravitational torque from the rest of Mercury ($e_s \alpha_3\alpha_g$) and the Sun ($\alpha_3 \phi_s$) acting on the inner core.  For both of our inner core density scenarios (and our choices of $\rho_s=8800$ kg m$^{-3}$ and $\alpha_3=0.1$), the internal  gravitational torque dominates that from the Sun.  Furthermore, $\alpha_3 \alpha_g\gg  \alpha_1$; the gravitational torque dominates the inertial torque, in large part because of the slow rotation rate of Mercury.  As a result the FICN frequency is negative (i.e. the precession motion is retrograde).  This is also the case for the Moon \cite[e.g.][]{dumberry16,stys18}, but it is different for Earth, where $\alpha_1 > \alpha_3 \alpha_g$ because of its faster rotation and the FICN mode is prograde \cite[][]{mathews91a}.  Note also that our approximate expression for the FICN differs by a factor $(\bar{A}+\bar{A}_s)/(\bar{A}-\bar{A}_s)$ compared to that given in \cite{dumberry16} and \cite{stys18} for the Moon.  

The expression for FCN frequency differs from the usual expression for Earth.  First, it involves the external torque from the Sun captured by the parameter $\phi_m$.  If we set $\phi_m=0$, we obtain the FCN frequency for a decoupled model in which only interior torques contribute, 

\begin{equation}
\omega_{fcn,int}  \approx  -\Omega_o \left(\frac{\bar{A}}{\bar{A}_m + \bar{A}_s} \right) e_f  \, . \label{eq:fcnint} 
\end{equation}
\label{eq:fcnficn} 
\end{subequations}
This frequency is slightly different from the usual expression for Earth, involving the ratio $\bar{A}/(\bar{A}_m + \bar{A}_s)$ rather than $\bar{A}/\bar{A}_m$.  This is because of the relatively thin mantle of Mercury;  for the largest $r_s$ considered, the moment of inertia of the inner core can get close to 40\% of that of the mantle and is not negligible.  The period of the FCN when only interior torques contribute is shown in Figure \ref{fig:dens}b. It is close to 1100 yr  for a small inner core, increasing to approximately $1500$ yr at the largest $r_s$.  Hence, the influence of the solar torque reduces the FCN period by a factor of approximately 3. We note that the FICN period, in contrast, is not altered substantially when the external torque is set to zero.

\subsection{Gravitational and inertial coupling}

\begin{figure}
\begin{center}
\centerline{ \includegraphics[height=5.5cm]{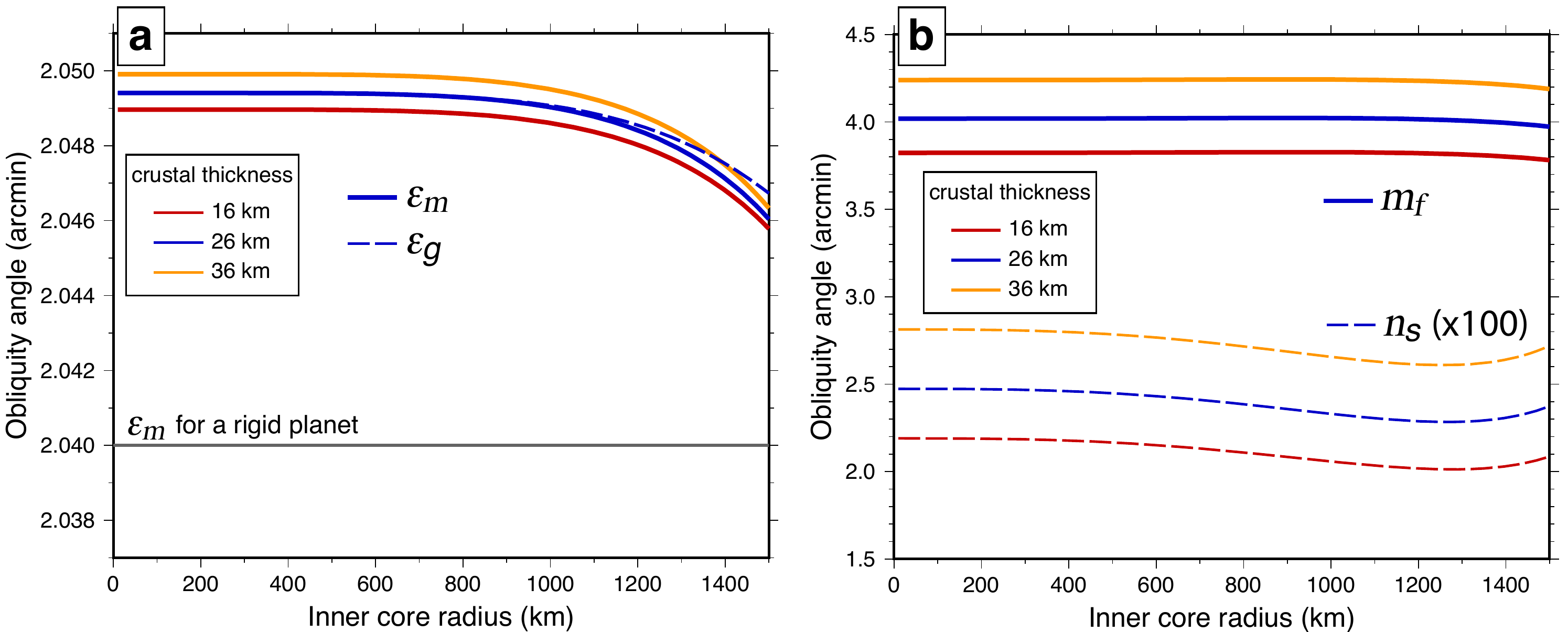} }
\caption{\label{fig:obl} a) Obliquity of the mantle ($\tilde{\varepsilon}_m$, solid lines) and of the principal moment of inertia ($\tilde{\varepsilon}_g$, dashed line) b) $\tilde{m}_f$ (solid lines) and $\tilde{n}_s$ (dashed lines, x100) as a function of inner core radius and for different choices of crustal thickness.}
\end{center}
\end{figure}

Let us now investigate the obliquities of the mantle, fluid core and inner core in their equilibrium Cassini state.  We assume a fixed inner core density scenario in this section, with $\rho_s=8800$ kg m$^{-3}$.  Viscous and EM coupling are set to zero in order to isolate the influence of gravitational and inertial coupling. Figure \ref{fig:obl} shows how $\tilde{\varepsilon}_m$, $\tilde{m}_f$ and $\tilde{n}_s$ vary as functions of inner core radius.  We show calculations for three different choices of crustal thickness, but let us concentrate first on the case for $h=26$ km.  For small $r_s$, we retrieve an obliquity of $\tilde{\varepsilon}_m=2.0494$ arcmin (Figure \ref{fig:obl}a).    $\tilde{\varepsilon}_m$ decreases with $r_s$, but not substantially; at the largest $r_s$ ($1500$ km),   $\tilde{\varepsilon}_m=2.0460$ arcmin, a decrease of 0.0034 arcmin.   The maximum difference from $\tilde{\varepsilon}_m=2.04$ arcmin, the obliquity that we used in setting the constraint for $\hat{C}$ -- and hence the prediction we should recover for a rigid planet -- is an overestimate of approximately $0.01$ arcmin which occurs for small inner cores.   

The deviation of $\tilde{\varepsilon}_m$ from that of a rigid planet is due to the misalignments of the fluid core ($\tilde{m}_f$) and solid inner core ($\tilde{n}_s$) with respect to the mantle (Figure \ref{fig:obl}b).  The misalignment of the fluid core spin axis from the mantle is significant: $\tilde{m}_f$ is approximately 4.02 arcmin for a small inner core and does not vary substantially with inner core size; it drops to 3.97 arcmin at the largest $r_s$. Recall that $\tilde{m}_f$ is measured with respect to the mantle rotation axis (which coincides closely with the symmetry axis), so the obliquity of the spin axis of the fluid core with respect to the orbit normal is $\tilde{\varepsilon}_m+\tilde{m}_f\approx 6$ arcmin.  The reason why the  obliquity of the spin axis of the fluid core is larger than that of the mantle can be understood from Equation (\ref{eq:approxmf}), which shows that $\tilde{m}_f$ is determined by the resonant amplification of the FCN mode at the forcing frequency.  When the FCN frequency is much larger than the forcing frequency, as is the case for Mercury, the resonant amplification is very weak but remains present and $\tilde{m}_f$ is larger than zero.

In contrast to $\tilde{m}_f$, the misalignment of the inner core with respect to the mantle is much smaller; $\tilde{n}_s$ is approximately between 0.023-0.025 arcmin, a factor 80 times smaller than $\tilde{\varepsilon}_m$.  Physically, this is because the gravitational torque acting on the inner core when it is tilted from the mantle is much stronger than the inertial torque acting at the ICB. As a result, the inner core must remain in close alignment with the mantle.  Presented differently, since the FICN period is more than 3000 times shorter than the forced precession period, the inner core can easily follow the forced precession of the mantle and remains gravitationally locked to it.  $\tilde{n}_s$ does not change substantially as the inner core increases in size. 

When $K_{icb}=K_{cmb}=0$, a good approximation of $\tilde{\varepsilon}_m$ is given by

\begin{equation}
\tilde{\varepsilon}_m = \frac{{C'} \Omega_p \sin I}{-{C'} \Omega_p \cos I +  \bar{A} \Omega_o \phi_m } \, ,
\label{eq:predepsmK0}
\end{equation}
which is identical to the prediction of Equation (\ref{eq:predepsm1}) for a rigid Mercury, except $C$ is replaced by $C'$.  The latter represents an effective moment of inertia that accounts for the coupling of the core to the mantle,

\begin{equation}
C' = C + \bar{A}_c \chi  \, ,
\label{eq:Cprime}
\end{equation}
where  $\bar{A}_c=\bar{A}_f+\bar{A}_s$ and 

\begin{equation}
 \chi  = \frac{\Omega_p \cos I}{\bar{A}_c} \left( \frac{\bar{A}_f}{(\bar{\sigma}_f - \Omega_p \cos I)} + \frac{\bar{A}_s}{(\bar{\sigma}_s - \Omega_p \cos I)} \right)   - \frac{\bar{A}_s}{\bar{A}_c} \frac{\Omega_o \alpha_3\phi_s}{(\bar{\sigma}_s - \Omega_p \cos I)} \, .
\label{eq:chiK0}
\end{equation}
The frequencies $\bar{\sigma}_f$ and $\bar{\sigma}_s$ are given in Equations (\ref{eq:sigmaf}-\ref{eq:sigmas})  and closely approximate the FCN and FICN frequencies of Equations (\ref{eq:fcnint}) and (\ref{eq:ficn}), respectively.  The factor $\chi$ captures then how the core is entrained to precess with the mantle, with the coupling between the two expressed in terms of the resonant amplification of the FCN and FICN frequencies.   In the limit of $\bar{\sigma}_f, \bar{\sigma}_s \rightarrow 0$, then $\chi=-1$, $C'=C_m$, the core is fully decoupled from the mantle and we retrieve Equation (\ref{eq:angmommantle}).  If instead $\bar{\sigma}_f, \bar{\sigma}_s \rightarrow \infty$, then $\chi=0$, $C'=C$ and we retrieve the prediction for a rigid planet. When both the FCN and FICN frequencies are much larger than $\Omega_p$, as is the case here, resonant amplification is weak, $\chi$ is small and positive, $C'>C$ and this leads to a slightly larger $\tilde{\varepsilon}_m$ compared to a rigid planet.  Because the inner core core is gravitationally locked to the mantle, deviations from a rigid planet are dominantly caused by the misalignment of the fluid core.   In Equation (\ref{eq:chiK0}),  $\bar{\sigma}_s \gg\bar{\sigma}_f$, so to a good approximation 

\begin{equation}
 \chi  \approx \frac{\bar{A}_f}{\bar{A}_c} \frac{\Omega_o \cos I}{(\bar{\sigma}_f - \Omega_p \cos I)}  \, .
\label{eq:chiK02}
\end{equation}
For a small inner core, $\chi\approx7.55\times10^{-3}$.  As the inner core grows, $\bar{A}_f$ decreases, and the combination $\bar{A}_c \chi$ also decreases.  This implies that $C'$ decreases with inner core size and, consequently, $\tilde{\varepsilon}_m$ also decreases with inner core size, as seen in Figure \ref{fig:obl}a, though it remains larger than the prediction for a rigid planet.

The specific predictions of  $\tilde{\varepsilon}_m$, $\tilde{m}_f$ and $\tilde{n}_s$ on Figure \ref{fig:obl} depend sensitively on the assumed interior density model and on the dynamical ellipticities of the inner core ($e_s$) and fluid core ($e_f$).  Hence, it depends on the choices we have made for the inner core density $\rho_s$, the crustal density $\rho_c$ and its thickness $h$.  Changing $\rho_s$, $\rho_c$ and/or $h$ requires a different combination of $\rho_f$, $\rho_m$ and $r_f$ in order to match $M$, $\hat{C}$ and $\hat{C}_m$.  In turn, this leads to different ellipticities at interior boundary in order to match $J_2$ and $C_{22}$, and thus different predictions for $\tilde{\varepsilon}_m$, $\tilde{m}_f$ and $\tilde{n}_s$.  To illustrate this, we show on Figure \ref{fig:obl} two additional predictions computed with crustal thicknesses changed to $h=16$ and 36 km.  The change in $\tilde{\varepsilon}_m$ remains modest,  $\sim 0.025\%$, but the changes in $\tilde{m}_f$ and $\tilde{n}_s$ are more substantial, $\sim5\%$ and $\sim10\%$, respectively. 

We also show on Figure  \ref{fig:obl}a (only for $h=26$ km) the obliquity of the principal moment of inertia of the whole planet, which we denote by $\tilde{\varepsilon}_g$.  A difference between $\tilde{\varepsilon}_g$ and  $\tilde{\varepsilon}_m$ occurs if the inner core is misaligned with the mantle.  As seen in the mantle frame, a tilted inner core (with $\tilde{n}_s$ assumed small) leads to an off-diagonal component of the moment of inertia tensor of $(C_s-\bar{A}_s) \alpha_3 \tilde{n}_s = \bar{A}_s e_s \alpha_3 \tilde{n}_s$.  The angle by which the mantle frame must be rotated so that the moment of inertia of the whole planet is purely diagonal is $(\bar{A}_s e_s \alpha_3 \tilde{n}_s)/(\bar{A} e)$, and hence a good approximation of $\tilde{\varepsilon}_g$ is

\begin{equation}
\tilde{\varepsilon}_g = \tilde{\varepsilon}_m + \frac{\bar{A}_s e_s}{\bar{A} e} \alpha_3 \tilde{n}_s \, .
\end{equation}
Since the inner core is gravitationally forced into a close alignment with the mantle, the difference between $\tilde{\varepsilon}_g$ and $\tilde{\varepsilon}_m$ remains very small.  For the largest inner core radius that we have considered, $\tilde{\varepsilon}_g$ differs from $\tilde{\varepsilon}_m$ only by approximately 0.001 arcmin. 

\subsection{Viscous coupling}

We now investigate how viscous coupling at the CMB and ICB affects the equilibrium Cassini state. \cite{peale14} present two different parameterizations of viscous coupling based on the timescale of attenuation of the differential rotation between the fluid core and mantle.  More complete analytical solutions for the flow resulting from a differentially precessing shell have been derived \cite[e.g.][]{stewartson63,busse68,rochester76} and we exploit these solutions here.  The parametrization of the viscous coupling constants $K_{cmb}$ and $K_{icb}$ based on them are given in \cite{mathews05}, 

\begin{subequations}
 \begin{align}
 K_{cmb} &= \frac{\pi \rho_f r_f^4}{\bar{A}_f} \sqrt{\frac{\nu}{2\Omega_o}} \Big( 0.195 - 1.976 i \Big) \, ,
 \label{eq:Klamcmb} \\
 K_{icb} &= \frac{\pi \rho_f r_s^4}{\bar{A}_s} \sqrt{\frac{\nu}{2\Omega_o}} \Big( 0.195 - 1.976 i \Big) \, ,
 \label{eq:Klamicb}
 \end{align}
 \label{eq:Klam}
 \end{subequations}
where $\nu$ is the kinematic viscosity.  The appropriate numerical value for $\nu$ in planetary interior is not well known but based on theoretical and experimental studies it is expected to be of the order of $10^{-6}$ m$^2$ s$^{-1}$ \cite[e.g.][]{gans72,dewijs98,alfe00,rutter02b,rutter02a}.  

The above parameterizations are valid only under the assumption that the flow in the boundary layer remains laminar.  Whether this is reasonable can be assessed by evaluating the Reynolds number $Re = r_f \Delta u_{f} /\nu$, associated with the differential velocity $\Delta u_{f}=r_f \Omega_o \tilde{m}_f$ at the CMB.  For $r_f=2000$ km, and taking $\tilde{m}_f=4$ arcmin $\approx 0.001$ rad from the results in the previous section, we get $\Delta u_{f} \sim 2$ mm/s and $Re\sim 6 \times 10^9$. Such a large Reynolds number indicates that the viscous friction between the fluid core and mantle should induce turbulent flows, as is the case for the Cassini state of the Moon \cite[][]{yoder81,williams01,cebron19}. For a boundary layer that involves turbulent flows, the viscous torque should be independent of the fluid viscosity and proportional to the square of the differential velocity.  The coupling constant $K_{cmb}$ should be in the form 

\begin{equation}
K_{cmb} =  f_{cmb} \big|  \tilde{m}_f \big| \Big( 0.195 - 1.976 i \Big) \, ,
\end{equation}
where $f_{cmb}$ is a numerical factor that depends among other things on surface roughness.  Incorporating a viscous coupling of this form in our rotational model is more challenging not only because $f_{cmb}$ is not known but also because the viscous torque is no longer linear in $\tilde{m}_f$.  One strategy is to find solutions through an iterative process.  The simpler alternative strategy that we adopt is to use the laminar formulas of Equation (\ref{eq:Klam}) but with the understanding that $\nu$ represents an effective turbulent viscosity.  

To give an estimate of an appropriate turbulent value for $\nu$, we turn to the Cassini state of the Moon.  A measure of the viscous dissipation at the CMB of the Moon has been obtained by fitting a rotation model to the librations of the Moon observed by Lunar Laser Ranging (LLR) \cite[][]{williams01,williams14,williams15}.  Viscous dissipation is reported in terms of a coupling parameter ${\cal K}$ and a recent estimate is ${\cal K}/C_L = (1.41 \pm 0.34) \times10^{-8}$ day$^{-1}$ \cite[][]{williams15}, where $C_L$ is the lunar polar moment of inertia. The connection between ${\cal K}$ and $K_{cmb}$ is

\begin{equation}
\Big| Im[K_{cmb}] \Big| = \frac{{\cal K}}{C_L} \frac{C_L}{C_{fL}} \frac{1}{\Omega_L} \, ,
\end{equation}
where $C_{fL}$ is the moment of inertia of the lunar core and $\Omega_L=2.66 \times 10^{-6}$ s$^{-1}$ the lunar rotation rate.  With $C_{fL}/C_L \sim 7 \times 10^{-4}$ \cite[e.g.][]{williams14}, this gives $| Im[K_{cmb}] |\sim 9 \times 10^{-5}$.  In order to match this amplitude in Equation (\ref{eq:Klamcmb}), with lunar parameters and assuming a lunar core radius of 400 km, the required turbulent viscosity is $\nu \approx 5 \times 10^{-4}$ m$^2$ s$^{-1}$, about 500 times larger than the laminar viscosity.  Note that the differential velocity at the CMB of the Moon is closer to 3 cm/s \cite[][]{yoder81,williams01}, more than 10 times larger than our estimate for Mercury above.  Since the effective turbulent coupling constant $K_{cmb}$ is proportional to the differential velocity, the effective turbulent viscosity appropriate for Mercury should be smaller.  Thus, $\nu \approx 5 \times 10^{-4}$ m$^2$ s$^{-1}$ gives a conservative upper bound for the possible effective turbulent viscosity that can be expected for Mercury.    

\begin{figure}
\begin{center}
\centerline{  \includegraphics[height=6.5cm]{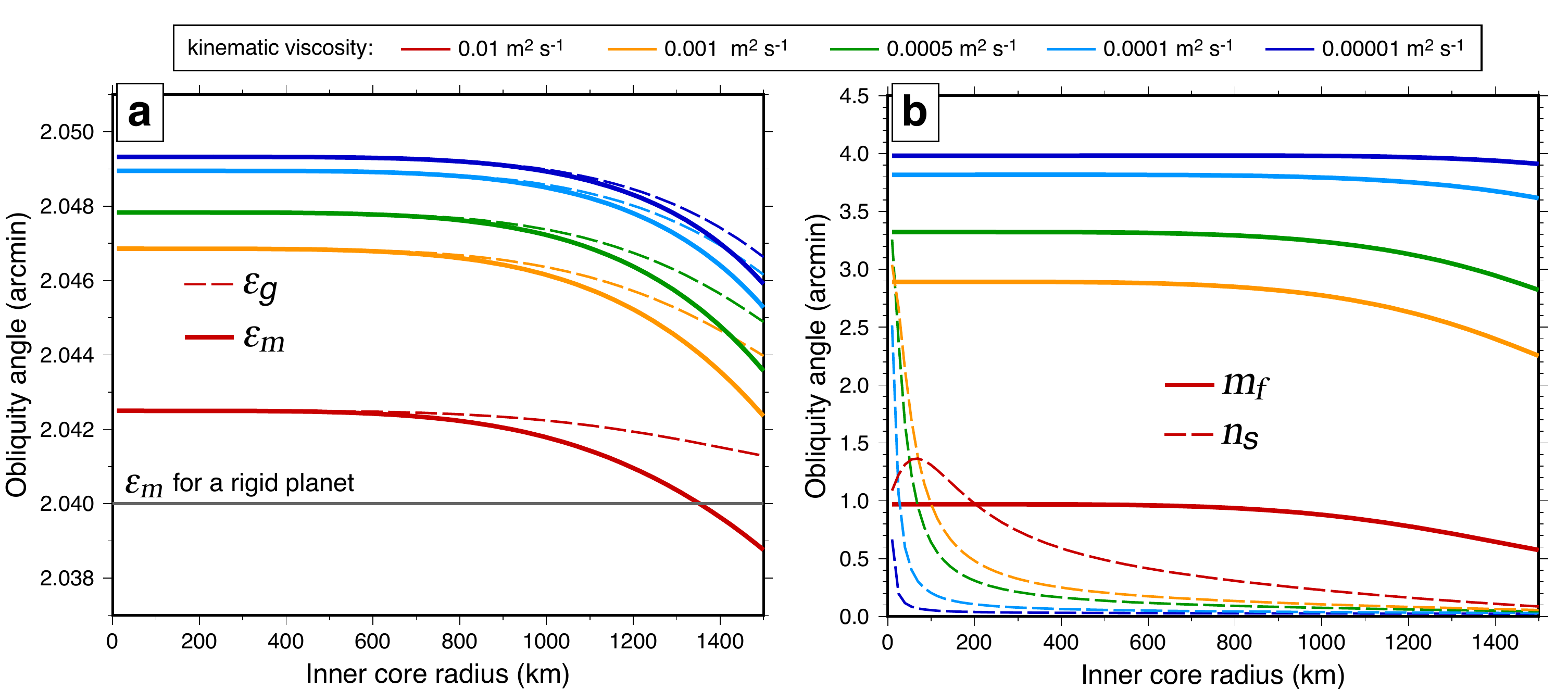} }
\caption{\label{fig:oblnu} a) Obliquity of the mantle ($\tilde{\varepsilon}_m$, solid lines) and gravity field ($\tilde{\varepsilon}_g$, dashed lines) b) $\tilde{m}_f$ (solid lines) and $\tilde{n}_s$ (dashed lines) as a function of inner core radius and for different choices of kinematic viscosity (color in legend).}
\end{center}
\end{figure}

Figure \ref{fig:oblnu} shows how $\tilde{\varepsilon}_m$, $\tilde{m}_f$ and $\tilde{n}_s$ vary as functions of inner core radius for different choices of effective viscosities. For $\nu = 10^{-5}$ m$^2$ s$^{-1}$, viscous coupling is too weak to affect $\tilde{\varepsilon}_m$ and $\tilde{m}_f$ and they are essentially unchanged from the solutions shown in Figure \ref{fig:obl}.  With increasing $\nu$, the stronger viscous coupling between the core and the mantle reduces their differential velocity, and $\tilde{m}_f$ is reduced. With the reduced differential velocity at the CMB, the prediction of $\tilde{\varepsilon}_m$ gets closer to 2.04 arcmin, the obliquity expected for a rigid planet.   Although our CMB viscous coupling model is different than the one used by \cite{peale14}, our results for  $\tilde{\varepsilon}_m$ and $\tilde{m}_f$ are qualitatively similar: viscous coupling at the CMB acts to reduce the offset of the fluid spin axis from the mantle symmetry axis.   Considering the upper bound in turbulent viscosity that we have identified above (i.e $\nu\approx5\times10^{-4}$ m$^2$ s$^{-1}$), the influence of viscous coupling on $\tilde{\varepsilon}_m$ remains modest, reducing its amplitude by a maximum of approximately 0.0015 arcmin.  

The inclusion of viscous coupling at the ICB can lead to a substantial change in inner core tilt. A larger viscosity leads to stronger viscous coupling and to a closer alignment of the inner core with the fluid core spin axis.  The viscous coupling strength is inversely proportional to $r_s$, so a larger viscosity results in a larger inner core radius at which viscous coupling is of a similar magnitude to gravitational coupling.  Taking again an upper bound of $\nu=5\times 10^{-4}$ m$^2$ s$^{-1}$, Figure \ref{fig:oblnu} indicates that $\tilde{n}_s$ may be 1 arcmin or larger only if the inner core radius is smaller than approximately 100 km.  For an inner core of a few hundred km in radius, gravitational coupling is much larger than viscous coupling, and the inner core tilt is limited to a fraction of 1 arcmin.   

The larger inner core tilt observed with increasing effective viscosity results in a larger offset between the obliquity of the principal moment of inertia $\tilde{\varepsilon}_g$ and that of the mantle $\tilde{\varepsilon}_m$, though it remains limited.  For the upper bound of $\nu=5\times 10^{-4}$ m$^2$ s$^{-1}$, and for $r_s=1500$ km, the difference between $\tilde{\varepsilon}_g$ and $\tilde{\varepsilon}_m$ is limited to 0.0013 arcmin.

The conclusion that emerges from Figure \ref{fig:oblnu} is that the larger the inner core is, the smaller the misalignments of both the fluid core and inner core are with respect to the mantle.  This implies that the larger the inner core is, the more we approach a planet precessing as a rigid body, although the misalignment of the spin axis of the fluid core remains important, approximately 3-4 arcmin away from the mantle symmetry axis. The specific way in which  $\tilde{\varepsilon}_m$, $\tilde{m}_f$ and $\tilde{n}_s$ change with inner core size would certainly be different for a turbulent model of viscous coupling.  But the general conclusion remains that the addition of viscous coupling at the CMB and ICB does not significantly modify the Cassini state equilibrium angle of the mantle.

\subsection{Electromagnetic coupling}
  
Let us now turn to electromagnetic (EM) coupling.  To focus on its role in the equilibrium Cassini state, we set the viscous coupling back to zero.  Because magnetic field lines tend to remain attached to electrically conducting materials, a differential tangential motion between two electrically conducting regions stretches existing magnetic field lines that thread their interface. This  induces a secondary magnetic field (or equivalently, an electrical current) and an associated tangential EM stress resisting the differential motion.  EM coupling at the CMB and ICB acts then in a similar way to viscous coupling, and this 'magnetic friction' depends on the strength of the radial magnetic field $B_r$ and the electrical conductivity $\sigma$ on either side of the boundary \cite[][]{rochester60,rochester62,rochester68}.

The parametrization of EM coupling in terms of the coupling constants $K_{cmb}$ and $K_{icb}$ has been developed in a few studies  \cite[e.g.][]{buffett92,buffett02,dumberry12}.   Assuming a dominating axial dipole field, with a radial component at the CMB given by $B_r = \sqrt{3} \left<B_r^d \right>\cos \theta$, where $\left<B_r^d\right>$ is the r.m.s. strength of the field, the coupling constant $K_{cmb}$ can be written is the form 

 \begin{equation}
 K_{cmb} = 3(1-i){\cal F}_{cmb} \left<B_r^d \right>^2 \, ,\label{eq:Kem}
 \end{equation}
 where 
 
 \begin{equation}
{\cal F}_{cmb} = \frac{1}{\Omega_o \rho_f r_f} \left( \frac{1}{\sigma_m \delta_m} + \frac{1}{\sigma_f \delta_f} \right)^{-1}\, , \label{eq:Fcmb}
 \end{equation}
and where $\sigma_m$, $\delta_m=\sqrt{2/(\sigma_m \mu \Omega_o)}$ and $\sigma_f$, $\delta_f=\sqrt{2/(\sigma_f \mu \Omega_o)}$ are the electrical conductivities and magnetic skin depths in the mantle and fluid core, respectively, with $\mu=4\pi \times10^{-7}$ N A$^{-2}$ the magnetic permeability of free space. The r.m.s. field strength $\left<B_r^d\right>$ is connected to the Gauss coefficient $g_1^0$ of the surface magnetic field by 

\begin{equation}
\left<B_r^d\right> = \frac{2}{\sqrt{3}} \left( \frac{R}{r_f} \right)^3 \left| g_1^0 \right| \, .
\end{equation}

We can readily build an estimate of the amplitude of $K_{cmb}$.  The electrical conductivity of common mantle minerals in Earth's mantle at the pressure and temperature corresponding to the CMB of Mercury is in the range of $\sigma_m \sim 0.01-1$ S m$^{-1}$ \cite[][]{constableTOG15}.   In contrast, the electrical conductivity of Fe in planetary cores is expected to be close $\sigma_f \sim 10^6$ S m$^{-1}$ \cite[][]{pozzo12,dekoker12}.  This implies that $(\sigma_m \delta_m)^{-1} \gg (\sigma_f \delta_f)^{-1}$.  Taking  $\sigma_m = 1$ S m$^{-1}$, $\left| g_1^0 \right|=190$ nT for Mercury's dipole field \cite[][]{anderson12}, $r_f=2000$ km, $\rho_f=7000$ kg m$^{-3}$, this gives $K_{cmb} \approx (3.1 \times 10^{-11})\cdot(1-i)$.  To put this amplitude in perspective, taking a molecular viscosity of $\nu = 10^{-6}$ m$^2$ s$^{-1}$ in Equation (\ref{eq:Klamcmb}) gives a viscous coupling constant of $K_{cmb} \approx  (6.0 \times 10^{-7}) \cdot ( 0.195 - 1.976 i )$.  Hence, EM coupling at the CMB is much weaker than viscous coupling, even if we include other spherical harmonic components of the radial magnetic field.  

EM coupling can be enhanced if strongly stratified pockets of core fluid are trapped by CMB cavities \cite[][]{buffett10b,glane18}, in which case the effective $\sigma_m$ could be closer to $\sigma_f$.  Likewise, $\sigma_m$ can be increased if a more electrically conducting layer has formed at the bottom of Mercury's mantle, for instance by the upward sedimentation and compaction of solid FeS crystals precipitating out of the fluid core \cite[e.g.][]{hauck13}.  However, even in the extreme case of $\sigma_m=\sigma_f = 10^6$ S m$^{-1}$, $K_{cmb}\approx(1.6 \times 10^{-8}) \cdot(1-i) $, which remains smaller by a factor $\sim60$ than the smallest possible viscous coupling constant.  Viscous forces dominate the tangential stress on the CMB of Mercury.    

At the ICB, because we can expect the electrical conductivity in both the solid inner core and fluid core to be similar, and because the radial magnetic field is likely much stronger, EM coupling can be much larger and dominate viscous coupling.  We assume that the magnetic field morphology at the ICB is dominantly comprised of small spatial scales for example as predicted by the dynamo model of \cite{christensen06b}.  EM coupling in this case can be parametrized in terms of an equivalent uniform radial magnetic field $\left<B_r\right>$ capturing its r.m.s. strength \cite[][]{buffett02,dumberry12}.  Assuming an electrical conductivity $\sigma$ equal in the fluid and solid core, the coupling constant $K_{icb}$ can be written in the form 
  
 \begin{equation}
 K_{icb} = \frac{5}{4}(1-i){\cal F}_{icb} \left<B_r\right>^2 \, ,\label{eq:Kem}
 \end{equation}
 where 
 
 \begin{equation}
{\cal F}_{icb} = \frac{\sigma \delta}{\Omega_o \rho_s r_s} \, , \label{eq:FM}
 \end{equation}
and where $\delta=\sqrt{2/(\sigma \mu \Omega_o)}$ is the magnetic skin depth. As   ${\cal F}_{icb}$ is inversely proportional to $r_s$,  $K_{icb}$ is inversely proportional to inner core size.  Note that computing the EM coupling based on the r.m.s. strength $\left<B_r\right>$ rather than a true field morphology tends to overestimate the strength of the coupling \cite[][]{koot13}.  However, since the strength of the radial magnetic field at the ICB of Mercury is largely unknown, imperfections of the EM coupling model are absorbed in the range of possible $\left<B_r\right>$ values.

The parametrization of Equation (\ref{eq:Kem}) is only valid in a 'weak field' regime \cite[][]{buffett02}, when the feedback from the Lorentz force on the flow in the fluid core can be neglected.  When $\left<B_r\right>$ is sufficiently large, this is no longer the case.  EM coupling then enters a 'strong field' regime \cite[][]{buffett02,dumberry12,koot13} in which $K_{icb}$  increases linearly with $\left<B_r\right>$ instead of quadratically.  A good approximation of $K_{icb}$ calculated for Earth can be extracted from Figure 6a of \cite{dumberry12},

 \begin{equation}
 K_{icb}^E = (0.175 - i 0.138) \left<B_r\right> \, ,
 \end{equation}
where $\left<B_r\right>$ is in units of Tesla.  The superscript $E$ emphasizes that the numerical factors are appropriate for the parameter values adopted for Earth in the computation of \cite{dumberry12}. To adapt these numerical factors to Mercury, we write,

 \begin{equation}
 K_{icb} = (0.175 - i 0.138) \frac{{\cal F}_{icb}}{{\cal F}^E_{icb}} \left<B_r\right> \, ,
 \end{equation}
 where ${\cal F}^E_{icb}$ is defined as in Equation (\ref{eq:FM}) but using the parameters for Earth as defined in  \cite{dumberry12}.  These are $\Omega_o = 7.292 \times 10^{-5}$ s$^{-1}$, $\rho_s = 12846$ kg m$^{-3}$, $r_s=1221.5$ km, $\sigma=5\times10^5$ S m$^{-1}$, which gives ${\cal F}^E_{icb} = 90.36$ T$^{-2}$.  
 
To compute ${\cal F}_{icb}$, we assume an electrical conductivity of $\sigma=10^6$ S m$^{-1}$ in the core of Mercury \cite[e.g.][]{dekoker12,deng13}.  The transition between the weak and strong field regime occurs when $\left<B_r\right> \approx 1.53$ mT for the real part of $K_{icb}$.   $\left<B_r\right>$ at the ICB of Mercury is unknown.  The dynamo model of \cite{christensen06b} showed that the field geometry inside the core could be dominated by small length scales, yet only the weaker lower harmonics of the field would penetrate through a thermally stratified layer in the upper region of the fluid core and reach the surface. If so, the field strength inside the core can exceed the surface field strength by a factor 1000.  Taking a surface field strength equal to $\sim300$ nT \cite[e.g][]{anderson12}, $\left<B_r\right>$ at the ICB could be as large as 0.3 mT, corresponding to approximately 10\% of the field strength within Earth's core.  Given that it is perhaps unlikely that Mercury's field can be as high as that in Earth's core, in all likelihood EM coupling at the ICB of Mercury remains in the weak field regime.

\begin{figure}
\begin{center}
\centerline{  \includegraphics[height=6.5cm]{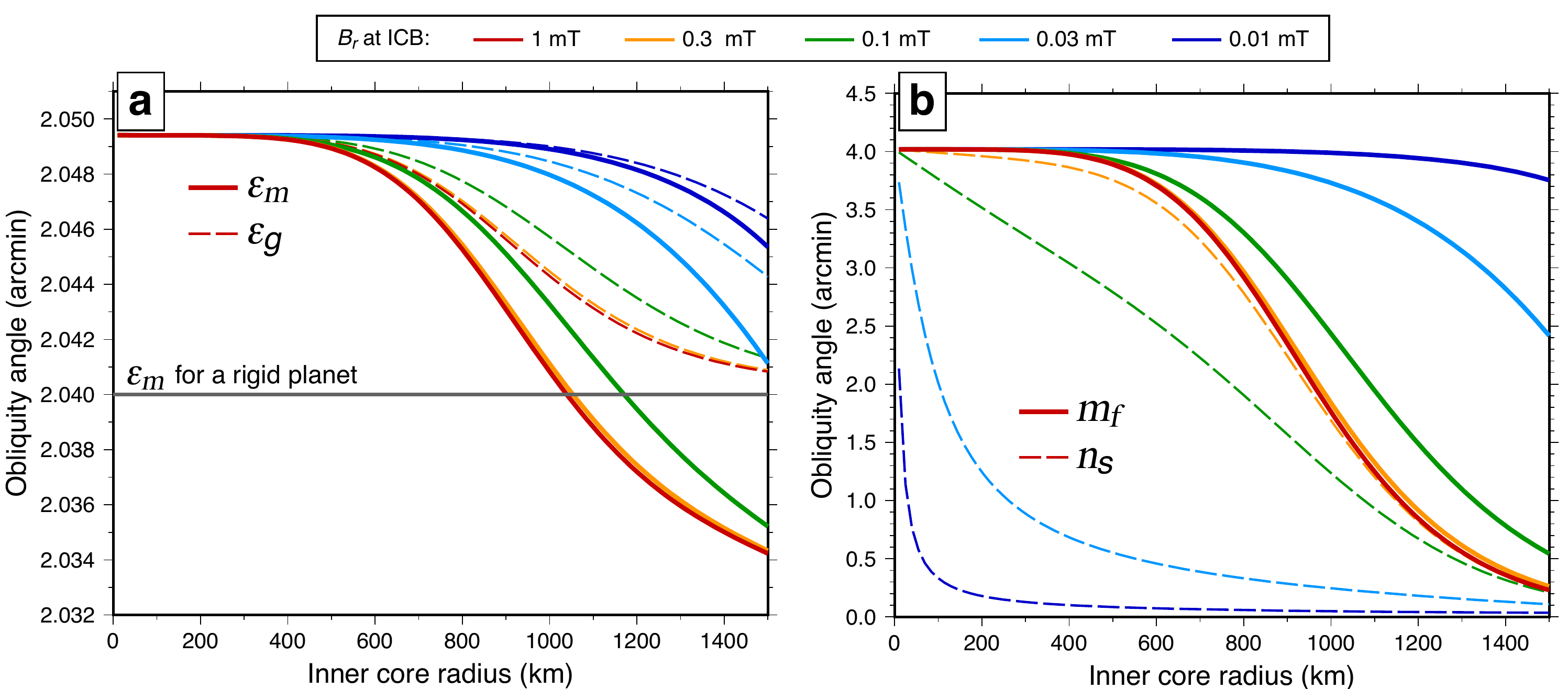} }
\caption{\label{fig:oblBr} a) Obliquity of the mantle ($\tilde{\varepsilon}_m$, solid lines) and gravity field ($\tilde{\varepsilon}_g$, dashed lines) b) $\tilde{m}_f$ (solid lines) and $\tilde{n}_s$ (dashed lines) as a function of inner core radius and for different choices of $B_r$ (colour in legend).}
\end{center}
\end{figure}

Figure \ref{fig:oblBr} shows how $\tilde{\varepsilon}_m$, $\tilde{m}_f$ and $\tilde{n}_s$ vary as functions of inner core radius for different choices of $\left<B_r\right>$.  The larger $\left<B_r\right>$ is, the stronger is the EM coupling at the ICB, and the smaller is the differential rotation between the fluid core and inner core.  The inner core and fluid core are virtually locked into a common precession motion when $\left<B_r\right> > 0.3$ mT.  Further increasing $\left<B_r\right>$ above 1 mT does not change the solution as EM coupling already dominates all other torques on the inner core.  This is the case even when EM coupling transitions into the strong field regime.  EM coupling at the CMB is included in these calculations, with $\sigma_m=1$ S m$^{-1}$ and $\left| g_1^0 \right|=190$ nT, but remains much weaker than the inertial torque at the CMB, so for a small inner core we retrieved the solutions of $\tilde{\varepsilon}_m$ and $\tilde{m}_f$ shown in Figure \ref{fig:obl}.  

As the inner core radius is increased, both $\tilde{\varepsilon}_m$ and $\tilde{m}_f$ get smaller, as it was the case with viscous coupling alone, although the addition of EM coupling lead to more substantial changes.  The inner core needs to be larger than approximately 500 km for changes in the Cassini state equilibrium to be noticeable.  It is important to point out that $\tilde{m}_f$ is reduced not because of EM coupling at the CMB, but rather from the combination of EM coupling at the ICB, which pulls the fluid core towards an alignment with the inner core, and gravitational coupling on the inner core, which pulls the latter to align with the mantle.  The larger the EM coupling is, the greater is the reduction in $\tilde{\varepsilon}_m$ and $\tilde{m}_f$.

When the EM coupling at the ICB is sufficiently strong that the fluid and solid cores are locked into a common precession motion, a good approximation of $\tilde{\varepsilon}_m$ is given by the same prediction as Equations (\ref{eq:predepsmK0}-\ref{eq:Cprime}) involving the effective moment of inertia $C'$, except $\chi$ is now given by

\begin{equation}
 \chi  = \frac{ \bar{A}_c \Omega_p \cos I - \bar{A}_s \Omega_o \alpha_3 \phi_s}{\bar{A}_f \Omega_o (e_f + K_{cmb}) + \bar{A}_s \Omega_o e_s \alpha_3 \alpha_g -  \bar{A}_c\Omega_p \cos I } \, .
\label{eq:chiKicbstrong}
\end{equation}
For a small inner core, $\bar{A}_c \Omega_p \cos I > \bar{A}_s\Omega_o \alpha_3 \phi_s$ and $\chi$ is positive. Because $\bar{A}_s\Omega_o \alpha_3 \phi_s$ increases with inner core size, $\chi$ gets smaller, and so do $C'$ and $\tilde{\varepsilon}_m$.   The mantle obliquity drops from 2.049 arcmin for a small inner core to 2.034 arcmin for an inner core of 1500 km, a reduction of 0.015 arcmin.    For an inner core larger than $\approx 1000$ km, $\bar{A}_c\Omega_p \cos I < \bar{A}_s\Omega_o \alpha_3 \phi_s$, so $\chi$ becomes negative, $C'$ becomes smaller than the moment of inertia of a rigid Mercury $C$, and $\tilde{\varepsilon}_m$ becomes smaller than the prediction based on a rigid planet.

The larger the inner core is, the smaller are the misalignments of the fluid and solid cores with respect to the mantle.   Hence, the general conclusion we reached for viscous coupling alone is not altered with the addition of EM coupling but further strengthened; the larger the inner core is, the closer we approach a planet precessing as a rigid body.  This is best revealed by the obliquity of the gravity field $\tilde{\varepsilon}_g$ which, for a large inner core, asymptotically approaches the obliquity expected for a rigid planet.  
Note that with strong EM coupling at the ICB, the offset between $\tilde{\varepsilon}_m$ and $\tilde{\varepsilon}_g$ can be as large as 0.008 arcmin for a large inner core. 

\subsection{Fixed inner core density versus fixed ICB density contrast}

Coupling models when viscous and EM stresses are both present have been presented in \cite{mathews05} and \cite{deleplace06}.  However, in the light of our results, for the Cassini state equilibrium of Mercury, the tangential stress at the CMB is dominated by viscous forces, and that at the ICB should be dominated by EM forces.  To simplify, we consider a model where $K_{cmb}$ is purely from viscous coupling and $K_{icb}$ purely from EM coupling.  We choose an effective viscosity at the CMB of $\nu= 10^{-4}$ m$^2$ s$^{-1}$, which we believe to be a representative value given the comparison with the Moon (see section 3.3).  We take a radial field strength at the ICB of $\left<B_r\right>=0.3$ mT, approximately the field strength expected under the dynamo scenario of \cite{christensen06b}.  We adopt these values as those of a `representative' coupling model, although the uncertainty on $\nu$ and $\left<B_r\right>$ obviously remains high.

\begin{figure}
\begin{center}
\centerline{  \includegraphics[height=6.5cm]{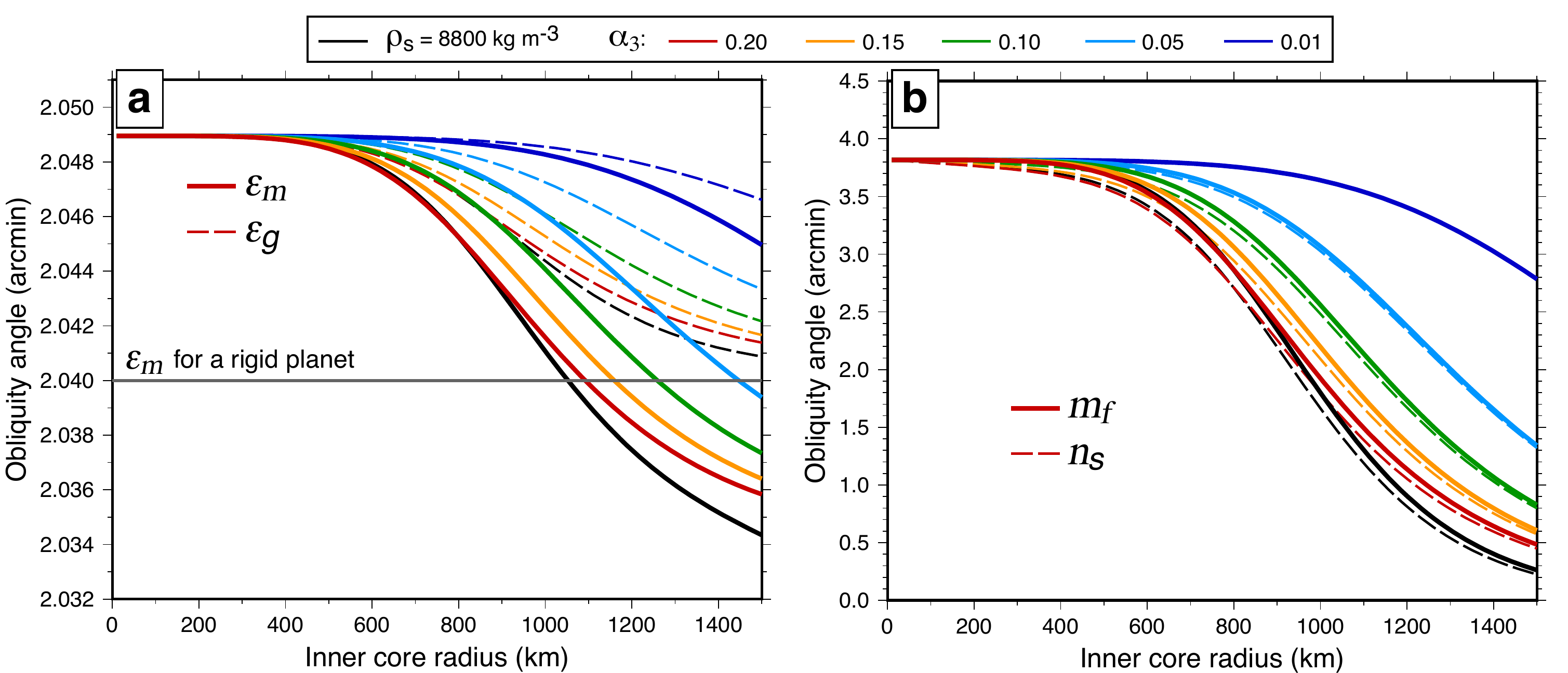} }
\caption{\label{fig:oblp} a) Obliquity of the mantle ($\tilde{\varepsilon}_m$, solid lines) and gravity field ($\tilde{\varepsilon}_g$, dashed lines) b) $\tilde{m}_f$ (solid lines) and $\tilde{n}_s$ (dashed lines) as a function of inner core radius, for a fixed inner core density of $8800$ kg m$^{-3}$ (black lines) and for different choices of $\alpha_3$ (coloured lines).}
\end{center}
\end{figure}

Figure \ref{fig:oblp} shows how $\tilde{\varepsilon}_m$, $\tilde{m}_f$ and $\tilde{n}_s$ vary with inner core radius for the 'representative' coupling model (black lines) under the fixed inner core density scenario that we have used in sections 3.2, 3.3 and 3.4.  Figure \ref{fig:oblp} also shows how the results change when, for the same representative coupling model,  we adopt instead a fixed density contrast between the fluid and solid cores and for different choices of $\alpha_3$ (coloured lines).  For a relatively high density contrast ($\alpha_3=0.2$), the results are qualitatively similar to the fixed inner core density scenario.  For a smaller $\alpha_3$, the point at which the orientation of the co-precessing fluid and inner cores begins to be pulled into an alignment with the mantle is pushed to a larger inner core radius.  However, the general behaviour of  $\tilde{\varepsilon}_m$, $\tilde{m}_f$ and $\tilde{n}_s$ as  functions of inner core radius is unchanged.  Hence, all our results in the previous three sections would be qualitatively similar under a fixed density contrast scenario.  A smaller density contrast at the ICB only implies that a larger inner core is required in order to produce an equivalent change in the Cassini state equilibrium.

\section{Discussion}

The study of \cite{peale16} also presented predictions of the obliquities of the mantle, fluid core and inner core associated with the equilibrium Cassini state of Mercury.  Their model included the tangential viscous stress at the ICB and CMB, but not the EM stress.  Their Table 1 gives the obliquities of the mantle, fluid core and inner core, denoted respectively as $i'_m$, $i'_f$ and $i'_s$; these represent the obliquities with respect to the orbital plane and are connected to our variables by: $i'_m = \tilde{\varepsilon}_m$, $i'_f = \tilde{\varepsilon}_m + \tilde{m} + \tilde{m}_f \approx \tilde{\varepsilon}_m + \tilde{m}_f$ and $i'_s = \tilde{\varepsilon}_m + \tilde{n}_s$.  To summarize their results, $i'_f$ and $i'_s$ vary substantially for different inner core sizes, are always of comparable amplitude, and $i'_s$ is always larger than $i'_f$.  Furthermore, they find that as the inner core size is increased, the mantle obliquity $i'_m$ gets progressively larger and is displaced further away from its expected orientation based of a rigid planet (see their Figure 6).  The change in $i'_m$ they obtain between a case with no inner core and an inner core radius equal to 0.6 times the planetary radius ($\approx1463$ km, close to the maximum inner core size of 1500 km we have considered), is approximately an increase of $5\times10^{-5}$ rad = 0.17 arcmin. This also corresponds approximately to the deviation of the obliquity with respect to that of a rigid planet.

When only viscous stress is included in our model (section 3.3), our results are substantially different.  As illustrated in Figure 4, we find instead that the obliquity of the fluid core gets smaller with inner core size and that the change is very modest.  In contrast with the results of \cite{peale16}, we find that the inner core obliquity is typically smaller than that of the fluid core, except when the inner core is very small or when the effective viscosity is unreasonably large. We also find that as the inner core size is increased, the mantle obliquity gets smaller, opposite to the results of \cite{peale16}, and that the changes remain small, at most of the order of 0.005 arcmin.  A part of the difference is due to the different viscous coupling model that we use.  But even when we adopt their model parameters and use their viscosity model, we were not able to reproduce their results.  

In the absence of viscous and EM coupling, the strong gravitational torque exerted on the inner core by the mantle should prevent any large misalignment between the two. This is captured by the period of the FICN, which is of the order of  100 yr, much shorter than the forcing period of 325 kyr.  Viscous and/or EM coupling at the ICB can counteract the gravitational torque (and alter the period of the FICN), but only for a small inner core.  The ratio of the viscous-EM torque to the gravitational torque decreases with inner core size, so a large inner core should be more strongly aligned with the mantle.  The more strongly the inner core and mantle are gravitationally locked together, the more they behave as a single rigid body in response to the external torque from the Sun.  We expect then that the obliquity of the mantle should be brought closer to that of a rigid planet when the inner core is larger.  Hence, we find puzzling the results of \cite{peale16}, which suggest the opposite.  

We showed that EM coupling is most likely larger than viscous coupling at the ICB, even though our knowledge of the radial magnetic field strength inside Mercury (on which EM coupling depends) remains poor.  If the magnetic field strength at the ICB is above 0.3 mT, EM coupling is sufficiently strong to bring the fluid and solid cores into a locked procession motion.  The larger the inner core is, the more this co-precessing core is forced into an alignment with the mantle because of the mantle gravitational torque on the inner core.   As a result, the larger the inner core is, the closer we approach a situation resembling a whole planet precessing as a rigid body.  The addition of EM coupling at the ICB does not change the overall picture that we observe with viscous coupling alone; the mantle obliquity decreases with inner core size.  The amplitude of the decrease can be as large as 0.015 arcmin, 3 times larger than for viscous coupling alone; this remains a factor 10 smaller than the changes suggested in \cite{peale16}, and again, importantly, in the reverse direction.  

Our results suggest then that the presence and size of an inner core leads to only modest changes of the mantle obliquity $\varepsilon_m$ compared to the obliquity predicted on the basis of an entirely rigid planet ($\varepsilon_m^r$).  Let us denote this difference as $\Delta \varepsilon_m = \varepsilon_m-\varepsilon_m^r$.  The largest $\Delta \varepsilon_m$ occurs for a small or no inner core, and is $\Delta \varepsilon_m\approx0.01$ arcmin.   This difference is decreased as the inner core size is increased.  For a sufficiently large inner core, in the case of a strong EM coupling and large density contrast at the ICB,  $\Delta \varepsilon_m$ can be negative, but its absolute value remains smaller than 0.01 arcmin.

To put these results in perspective,  the uncertainty in the measurement of the mantle obliquity reported by \cite{margot12} and \cite{stark15} is of the order of 0.08 arcmin, much larger than this difference. This means that, at the current level of precision, it is not possible to distinguish the position of the mantle obliquity from the obliquity of a rigid planet.  This is consistent with the fact that the observed obliquity falls close to that expected from a rigid planet. But it also implies that the observed obliquity cannot be used to place constraints on the inner core size.

Nevertheless, our results show that the presence of a fluid core and inner core affect the resulting mantle obliquity by as much as 0.01 arcmin.  This is of the same order as the change in obliquity caused by elastic tidal deformation, which is of the order of 0.35 arcsec ($\approx 0.006$ arcmin) \cite[][]{baland17}.  This is also of the same order as the amplitude of the nutation motion about the mean equilibrium Cassini state forced by the precession of the pericenter, which is approximately 0.85 arcsec ($\approx0.014$ arcmin) \cite[][]{baland17}.  The precision on the obliquity from the upcoming BepiColombo satellite mission is expected to be $\le0.5$ arcsec ($\le0.008$ arcmin) \cite[][]{cicalo16}.  Thus, in addition to including tidal deformation and the precession of the pericenter, a Cassini state model that includes a fluid and solid core will then be necessary in order to properly tie Mercury's obliquity to its interior structure.  In turn, this opens the possibility of further constraining the interior structure of Mercury on the basis of its obliquity.
 
Obliquity measurements based on tracking topographic features reflect the orientation of the spin-symmetry axis of the mantle ($\varepsilon_m$).  Measurements based on tracking the gravity field of Mercury reflect instead the orientation of the principal moment of the whole planet ($\varepsilon_g$). These two orientations do not coincide when an inner core is present and is misaligned from the mantle.  Since gravitational coupling prevents a large inner core tilt with respect to the mantle, we find that the misalignment $\Delta \varepsilon_g= \varepsilon_g -\varepsilon_m$ is limited.  The maximum offset that we obtain is approximately $\Delta \varepsilon_g\approx 0.007$ arcmin.  This limited magnitude of offset is important in the light of the recent obliquity of the gravity field estimated in \cite{genova19}, $\varepsilon_g=1.968\pm0.027$ arcmin.  This is substantially smaller than the two mesurements of the obliquity  of the spin-symmetry axis of the mantle: $\varepsilon_m=2.04\pm0.08$ arcmin \cite[][]{margot12} and $\varepsilon_m=2.029\pm0.085$ arcmin \cite[][]{stark15}, although all three measurements remain consistent with one another within their error estimates. In their interpretation, \cite{genova19} suggest that the different central value of the obliquity that they obtain (smaller by $\sim0.07$ arcmin) is perhaps explained by an offset $\Delta \varepsilon_g$ due to the presence of a (possibly large) solid inner core.  However, this is one order of magnitude larger than the maximum magnitude of $\Delta \varepsilon_g$ that we predict.  Moreover, we predict that the obliquity of the gravity field should be larger than that of the mantle spin axis, not smaller.  Hence, at the present-day level of the precision of the measurements, $\varepsilon_g$ and $\varepsilon_m$ should coincide, and their difference cannot be interpreted as reflecting the misalignment between the polar moment of inertia of the whole planet and the mantle spin axis.

Lastly, we have concentrated our efforts on the mutual orientations of the different spin and symmetry axes in the Cassini plane.  Dissipation at the CMB and ICB introduced by viscous and EM coupling also lead to a displacement of these axes in the direction perpendicular to the Cassini plane \cite[e.g][]{peale14}.   Indeed,  the two measurements based on tracking surface topographic features from \cite{margot12} and \cite{stark15} suggest that the mantle spin axis lags behind the Cassini plane by approximately 2 arcsec ($\sim0.03$ arcmin).  Although this offset is smaller than the measurement errors, so that the observed obliquity is still consistent with no deviation away from the Cassini plane, some amount of dissipation invariably takes place.  These measurements give then a measure of the possible amplitude of the dissipation.  One source of dissipation is from anelastic tidal deformation \cite[][]{baland17}, but viscous and EM coupling at the boundaries of the fluid core is another.  Hence, the out-of-plane component of the observed obliquity may further help to quantify and constrain the interior coupling mechanisms.  This will be the subject of a future study.

\section{Conclusion}

We have investigated how the presence of a fluid core and solid inner core affects the Cassini state equilibrium of Mercury.  Our general conclusion is that the coupling strength between Mercury's interior regions is sufficiently strong that the obliquity of the mantle spin-symmetry axis does not deviate from that of a rigid planet by more than 0.01 arcmin. This largest offset occurs for a small or no inner core.  The larger the inner core is, the more it is forced into an alignment with the mantle because of the strong gravitational torque between the two, and the closer we approach a situation resembling a whole planet precessing as a rigid body.  The misalignment between the polar moment of inertia and mantle spin axis increases with inner core size, but is limited to approximately 0.007 arcmin.  These conclusions apply irrespective of the core composition and thus of the partitioning of light elements into the solid core; a smaller density contrast at the ICB only implies that a larger inner core is required in order to produce an equivalent change in the Cassini state equilibrium.

Our results imply that the obliquities of the mantle spin axis and polar moment of inertia (or, equivalently, the gravity field) should coincide at the present-day level of measurement errors.  Moreover, neither of these can be distinguished from the obliquity predicted on the basis of a rigid planet.  However, the smaller measurement errors expected from the upcoming BepiColumbo satellite mission may permit this distinction, and thus provide further constraints on Mercury's interior structure.

 \acknowledgments
 Figures were created using the GMT software \cite[]{gmt}. The source codes, GMT scripts and data files to reproduce all figures are freely accessible in \cite{dumberry20}. This work was supported by an NSERC/CRSNG Discovery Grant.  
 

\end{document}